\documentclass[]{aastex631}

\shorttitle{Plausible Tenuous Atmospheres in the TRAPPIST-1 System}
\shortauthors{Gialluca, et al.}

\usepackage{amsmath,amssymb}
\usepackage{wrapfig}
\usepackage{tabularx}
\usepackage{xcolor}
\usepackage[table]{xcolor}
\usepackage{makecell}
\usepackage{multirow}
\usepackage{booktabs}
\usepackage{hhline}
\usepackage{nicematrix}
\usepackage{tikz}

\newcommand{\peerreview}[1]{#1}
\newcommand{\secondpeerreview}[1]{#1}

\begin{document}

\title{Coupled Photochemical-Climate Modeling of Plausible Tenuous Outgassed Atmospheres on the TRAPPIST-1 Planets}

\author[0000-0002-2587-0841]{Megan T. Gialluca}
\affiliation{Department of Astronomy and Astrobiology Program, University of Washington, Box 351580, Seattle, Washington 98195, USA}
\affiliation{SETI Institute, 339 Bernardo Ave, Suite 200, Mountain View, California, 94043, USA}

\author[0000-0002-1386-1710]{Victoria S. Meadows}
\affiliation{Department of Astronomy and Astrobiology Program, University of Washington, Box 351580, Seattle, Washington 98195, USA}
\affiliation{SETI Institute, 339 Bernardo Ave, Suite 200, Mountain View, California, 94043, USA}

\author[0000-0003-0429-9487]{Andrew P. Lincowski}
\affiliation{Eastern Wyoming College, 3200 W C Street, Torrington, Wyoming 82240, USA}
\affiliation{SETI Institute, 339 Bernardo Ave, Suite 200, Mountain View, California, 94043, USA}

\author[0000-0003-2457-2890]{Trent B. Thomas}
\affiliation{Department of Earth and Space Sciences and Astrobiology Program, University of Washington, Seattle, Washington 98195, USA}
\affiliation{SETI Institute, 339 Bernardo Ave, Suite 200, Mountain View, California, 94043, USA}

\author[0000-0001-9504-0520]{P.C. Hinton}
\affiliation{University of Colorado, Boulder, CO, USA}
\affiliation{Laboratory for Atmosphere and Space Physics, Boulder, CO, USA}

\author[0000-0001-8932-368X]{David Brain}
\affiliation{University of Colorado, Boulder, CO, USA}
\affiliation{Laboratory for Atmosphere and Space Physics, Boulder, CO, USA}

\author[0000-0002-4573-9998]{David Crisp}
\affiliation{Jet Propulsion Laboratory, California Institute of Technology, Earth and Space Sciences Division, Pasadena, CA 91011, USA}



\begin{abstract}

Available JWST observations TRAPPIST-1 system have suggested that several of the planets are likely airless, or possess a very tenuous atmosphere. However, the high atmospheric escape rates expected for these planets suggest that any tenuous atmosphere must be replenished by constant outgassing, and past studies on modeling potential atmospheres for the planets have not widely considered surface pressures $<<$1 bar. Here, we show that tenuous atmospheres on the TRAPPIST-1 planets are likely possible, supported by constant plausible rates of water and/or CO$_{2}$ outgassing against assumed high escape rates (up to $\sim$10$^{30}$ s$^{-1}$). 
We use a coupled photochemical-climate model and sample from a broad phase space of outgassing, surface deposition, and top-of-atmosphere escape rates to test hundreds of atmospheres per planet. Critically, our model also allows surface pressure to vary based on the balance of sources and sinks. We find that 6 different compositional archetypes are generated via H$_{2}$O and/or CO$_{2}$ outgassing across our phase space, and atmospheres commonly fall between 10$^{-4}$ -- 1 bar. We find that potentially habitable surface environments are possible for TRAPPIST-1d and e at pressures between 0.05 -- 2 bar and 0.5 -- 1 bar, respectively. 
Where possible, we compare our models to JWST observational data for TRAPPIST-1b, c, d, and e; all atmospheres found in this study for these planets match available transmission data to $<$3$\sigma$. However, emission data are consistent with atmospheric outcomes constrained to thin O$_{2}$-dominated compositions for TRAPPIST-1b ($\lesssim$0.01 bars) and c ($\lesssim$0.2 bars), which may or may not contain trace SO$_{2}$. 

\end{abstract}



\section{Introduction} \label{sec:intro}

The launch of JWST has enabled the first searches for high mean molecular weight atmospheres on Earth-sized terrestrial exoplanets. Though these observations are primarily confined to planets with M dwarf hosts \citep[][]{Morley2017jwstdetect,Lustig2019detectT1,wunderlich2019detectability,Gialluca2021characterizing,Gialluca2023corrigendum}, the harsh stellar environments of these systems provide a vital test case for our understanding of atmospheric retention under extreme conditions, especially when comparing to the known planetary evolution in the solar system with a G dwarf host star. In particular, the TRAPPIST-1 system has been the subject of intense scientific focus due to its observationally favorable conditions and the existence of seven Earth-sized planets, both within and outside of the habitable zone \citep[][]{Gillon2017seven}. \peerreview{Furthermore, the TRAPPIST-1 planets are only slightly less dense than those in the Solar System, which was revealed by precise transit timing variations \citep[][]{Agol2021ttv}.} The TRAPPIST-1 system provides an important test of our understanding of atmospheric escape and replenishment as a function of distance from the host star in the extreme environment of an M dwarf. 

Significant JWST time has already been invested in the TRAPPIST-1 \peerreview{(also abbreviated as T-1)} system \citep[][]{Greene2023t1b,Zieba2023t1cjwst,lim2023atmospheric,Howard2023T1flaringJWST,Piaulet2025t1d,Radica2025t1c,Ducrot2025combined}, but due to the intense stellar activity in transmission observations \citep[e.g.,][]{lim2023atmospheric}, constraints on the planetary atmospheres have primarily been \peerreview{inferred} from emission photometry on the interior two planets, T-1b and c \citep[][]{Greene2023t1b,Zieba2023t1cjwst,Ducrot2025combined}. 
\peerreview{Relatively high brightness temperatures, approaching the expected equilibrium dayside temperatures, have been found in emission for T-1b and c. This has led most previous studies \citep[e.g.,][]{Greene2023t1b,Zieba2023t1cjwst,ih2023constrainingb,Lincowski2023t1catms,Gillon2025phase} to conclude that the best fitting scenarios are either airless bodies or tenuous atmospheres, which would currently be observationally indistinguishable at reasonable confidence ($\gtrsim$3$\sigma$). However, it has been questioned if it is possible at all to sustain thin atmospheres on M dwarf terrestrials against the intense expected atmospheric escape \citep[e.g.,][]{Mansfield2024gl486b,Radica2025t1c}.} 

The first observations for both T-1b and c were taken in the 15 $\mu$m bandpass, which ruled out substantial ($\gtrsim$ 1 bar) CO$_{2}$-dominated atmospheres in both cases \citep[][]{Greene2023t1b,Zieba2023t1cjwst}. \peerreview{Subsequent observations \citep[][]{Radica2025t1c,Ducrot2025combined,Gillon2025phase} have further disfavored high pressure atmospheres for T-1b and c}. Currently, bare rock compositions or thin ($\sim$0.1 bar) atmospheres with no strong absorbers at 15 $\mu$m provide equally good fits to the available data for the interior \peerreview{two} planets \citep[][]{Greene2023t1b,ih2023constrainingb,Zieba2023t1cjwst,Lincowski2023t1catms,Radica2025t1c}. 
Although a primordial atmosphere would be lost, and a tenuous atmosphere would have a very short \peerreview{expected} lifetime \peerreview{due to} the stellar activity \citep[e.g.,][]{Howard2023T1flaringJWST}, \peerreview{we propose that it may be possible for outgassing to replenish secondary atmospheres and provide a steady state}. 
\peerreview{Recently, \citet{Thomas2025outgassrates} used a geochemical model to place boundaries on volatile outgassing rates using theoretical upper limits on mantle water mass fraction informed by solar system geology \citep{Guimond2023mantleminerology}, and observational constraints on the lack of H$_{2}$-dominated atmospheres for TRAPPIST-1 \citep[e.g.,][]{deWit2018trappist1Observe,Wakeford2019trappist1g}. However, whether these outgassing rates can sustain a present-day atmosphere has not been tested.}  

Although the outer planets beyond the orbit of T-1c are too cold to be suited for emission observations, transmission observations have been executed or planned; but, most data relevant to atmospheric characterization remains inconclusive or unpublished at the time of this work -- with the exception of T-1d and e. In the case of T-1d, 2 transit observations with NIRSpec/PRISM have suggested strict constraints on certain atmospheric scenarios \citep[][]{Piaulet2025t1d}. Specifically, thick cloud-free atmospheres ($>$1 bar) composed of 100\% CH$_{4}$, H$_{2}$O, CO$_{2}$, or CO are ruled out to high ($>$95\%) confidence, though these results have a large dependency on the treatment of the Transit Light Source Effect (TLSE) \citep[][]{Piaulet2025t1d}. Several thin atmospheres (such as thin Mars-like CO$_{2}$ or thin N$_{2}$ dominated with minor contributions from IR-active species), as well as thick atmospheres with high altitude clouds (e.g., Earth-like water clouds, Venus-like H$_{2}$SO$_{4}$ clouds), remain consistent with observations \peerreview{of T-1d} \citep[][]{Piaulet2025t1d}. Additionally, O$_{2}$ dominated atmospheres with contributions from IR-active species have not been tested at this time. 

In the case of TRAPPIST-1e, 4 recently published NIRSpec/PRISM transmission observations have provided no strong atmospheric constraints due to significant stellar contamination \citep[][]{Espinoza2025t1e,Glidden2025t1e}. Venus and Mars-like CO$_{2}$ dominated atmospheres are weakly disfavored for T-1e at 2$\sigma$ confidence, however, atmospheres dominated by N$_{2}$ with contributions from a stronger IR absorber (e.g., CO$_{2}$, CH$_{4}$) remain unconstrained \citep[][]{Espinoza2025t1e,Glidden2025t1e}. Moreover, preliminary data from a multi-cycle JWST program for T-1e was also recently published by \citet{Allen2025t1ePrelimResults} who presented 3 additional transit observations with NIRSpec/PRISM. These additional transmission observations possessed comparable signal to those of \citet{Espinoza2025t1e,Glidden2025t1e}, and did not change the atmospheric interpretation or confidence in distinguishing a flat line scenario. \citet{Allen2025t1ePrelimResults} did, however, highlight the potential for CO to be a contributing absorber, which may be tested with additional future planned observations. 

Transmission observations for T-1b, c, f, and g have been used to constrain flaring frequency, starspots, and faculae, but did not provide strong constraints on the planetary atmospheres because of this noise \citep[][]{lim2023atmospheric,Howard2023T1flaringJWST}. Fortunately, new techniques have been developed to disentangle the planetary signal within transmission observations, relying on observing back-to-back occultations of two or more planets \citep[][]{Rathcke2025stellarcorrect}. This has led to a notable observing award for 15 transit observations of T-1e, a promising habitable zone planet (GO 9256\footnote{https://www.stsci.edu/jwst/science-execution/program-information?id=9256}). 

It is vital that the community prepare to analyze upcoming observations of the habitable zone planets by understanding what atmospheres may be theoretically probable, allowing us to readily rule out various plausible scenarios. Moreover, by working to constrain the most likely bare rock and atmosphere scenarios that fit the data on the interior planets, we may also provide useful information for comparative planetology approaches to understanding the outer planets in the system prior to their observations \citep[e.g.][]{Gialluca2024implications}. Also note, 500 hours of JWST Director's Discretionary Time (DDT) time was recently awarded to observing terrestrial planets within and interior to the habitable zones of M dwarf stars \citep[][]{Redfield2024DDTTime,DDTwebsite}. The tools we develop and lessons we learn about habitability and atmosphere maintenance on the TRAPPIST-1 planets will be directly applicable to the targets in the upcoming DDT program. 

Previous attempts to match JWST observations of TRAPPIST-1 planets have generated atmospheric models through self-consistent photochemical-climate calculations for a small number of \peerreview{selected compositional} cases \citep[e.g.,][]{Zieba2023t1cjwst,ih2023constrainingb,Lincowski2023t1catms}, rather than systematically exploring a broad parameter space. This approach to fitting data often involves fixing an atmospheric pressure and bulk composition, which also lacks self-consistency with plausible geochemical evolution. For example, an atmospheric model may select a fixed mixing ratio of oxygen \citep[e.g.,][]{Lincowski2023t1catms}, instead of outgassing more geochemically realistic species (e.g., H$_{2}$O, CO$_{2}$) which may later generate oxygen through photochemistry. 
\peerreview{Interpreting data through this method, or by using simple steady-state models, is an effective tool for quickly analyzing incoming data, and may also provide enhanced computational efficiency. However,} 
more in-depth studies that \peerreview{allow composition to change based on coupled photochemical and climatic effects are also} needed to generate plausible atmospheric models for the planets under consideration. This self-consistency \peerreview{allows for a better understanding of temperature structure and trace gas profiles, which can have a large effect on spectral features \citep[e.g.,][]{Meadows2023t1biosigs}. Furthermore, balancing outgassing and escape processes in atmospheric models allows the inference of interior processes from observational constraints on atmospheres and bulk planetary composition.}

In this work, we employ \peerreview{coupled photochemical-climate modeling} across a broad parameter space of outgassing, surface deposition, and top-of-atmosphere escape rates for all 7 TRAPPIST-1 planets. We show that there are a range of tenuous atmospheres that are plausibly maintained at present day, consistent with current JWST data and constraints on planetary interior evolution from solar system data \citep[][]{Thomas2025outgassrates}. We further provide the community with a database of all realistic atmospheric spectra generated in this work for use in analyzing future TRAPPIST-1 observations (found on GitHub\footnote{https://github.com/mgialluca/T1Atmospheres\_Gialluca2026}). 
For this work, we develop a wrapper code to efficiently parallelize calls of our atmospheric modeling suite (Section \ref{sec:methods}), and include a routine that allows surface pressure to vary in response to \peerreview{equilibrating atmospheric sources and sinks.} This modeling approach is novel in that we achieve self-consistency in both photochemical-climate and outgassing-escape balance, i.e., we are not forcing constraints on the atmospheric pressure, composition, or temperature.

To model a range of secondary outgassed atmospheres on the TRAPPIST-1 planets, we are informed by geochemical models that predict CO$_{2}$, H$_{2}$O and SO$_{2}$ as prominent gases released from terrestrial interiors, as a function of planetary composition and overlying atmospheric pressure \citep[][]{Gaillard2021outgassing}.
For our simulations, we use the water and CO$_{2}$ outgassing rate distributions of \citet{Thomas2025outgassrates} as input, along with an array of atmospheric sinks. We further explore the effect of trace SO$_{2}$ by testing the addition of 100 ppm, 0.1\%, or 1\% SO$_{2}$ in the bottom layer of all stable atmospheric models.  
Through this work, we explore types of atmospheres that are both observationally and theoretically plausible for the planets with available data (T-1b, c, d, and e), and we prepare for upcoming observations of planets in the system by providing a database of self-consistent spectra. 

\section{Methods} \label{sec:methods}

This work uses photochemical, climate, and radiative transfer models in a coupled framework to generate a range of terrestrial secondary atmospheres balanced by interior outgassing, surface deposition, and top-of-atmosphere (TOA) escape. 
In particular, we employ the \textit{Atmos} photochemical model \citep[\S \ref{subsec:AtmosMethods};][]{Arney2016pods,Meadows2018ProximaB,Lincowski2018trappist}, the Line-By-Line ABsorption Coefficients model \citep[\textit{LBLABC};][]{Meadows1996smart}, the Spectral Mapping Atmospheric Radiative Transfer Tool \citep[\textit{SMART}, \S \ref{subsec:SMARTMethods};][]{Meadows1996smart,Crisp1997smart}, and the \textit{VPL Climate} model \citep[\S \ref{subsec:ClimateMethods};][]{Robinson2018life,Lincowski2018trappist,Meadows2023t1biosigs}. Across our modeling pipeline, we use the \citet{Peacock2019trappistspectrum} stellar spectrum of TRAPPIST-1, which is visualized in Figure 2 of \citet{Meadows2023t1biosigs}. 

These models are called by a python wrapper developed for this work to iterate self-consistent atmospheres to convergence, enabling the exploration of a broad range of input boundary conditions (\S \ref{subsec:ModelFrameworkMethods}). We improve on past work by massively parallelizing atmosphere generation on HPC resources to explore a wide range of boundary conditions, and we add a routine to automatically update surface pressure in response to outgassing and escape. 
In this work, we consider atmospheres sustained by H$_{2}$O outgassing alone, or by mixed H$_{2}$O-CO$_{2}$ outgassing; we further consider adding 100 ppm -- 1\% of SO$_{2}$ at the surface level of stable atmospheres created with these possible outgassing sources. \peerreview{Finally,} we describe two methods for quantifying the fit of modeled atmospheric spectra to available data: the Chi-squared goodness of fit test (\S\ref{subsec:goodnessoffitMethods}) and the Pearson Correlation Coefficient (\S\ref{subsubsec:CorrelationCoeffMethods}). 

\subsection{The Atmos Photochemical Model} \label{subsec:AtmosMethods}

The \textit{Atmos} photochemical model \citep[][]{Kasting1979photochem,Arney2016pods} self-consistently accounts for the effects of incident radiation on atmospheric composition. Of particular relevance to the present study, \citet{Lincowski2018trappist} updated \textit{Atmos} to treat M dwarf hosts more realistically by including new cross sections and a higher resolution sampling grid in the UV; this update includes an extended H$_{2}$O cross section, which may be critical for H$_{2}$O photolysis when shielded by other species \citep[][]{Ranjan2020photochemistryH2O}. \textit{Atmos} has already been used in several studies to model both anoxic and extremely oxidized atmospheres in the TRAPPIST-1 system \citep[e.g.,][]{Lincowski2018trappist,Zieba2023t1cjwst,Lincowski2023t1catms,Meadows2023t1biosigs}. In depth descriptions of \textit{Atmos} can be found in \citet{Arney2016pods,Meadows2018ProximaB,Lincowski2018trappist}, but we provide a basic description here. 

For a given model, species and reaction lists are specified to \textit{Atmos}, which applies a vertical transport scheme (including molecular and eddy diffusion) to the atmosphere divided into 200 plane-parallel layers assumed to be in hydrostatic equilibrium. \peerreview{There are 251 chemical reactions in our network, 49 of which are photolysis reactions. We include 59 long-lived species, 7 short-lived species (calculated in equilibrium at each timestep), and 1 inert gas (N$_{2}$); all species \secondpeerreview{used in any chemical reaction} are composed of hydrogen, oxygen, carbon, nitrogen, and/or sulfur. \secondpeerreview{However note, we do not model nitrogen-bearing outgassing sources in this study (e.g., N$_{2}$, NH$_{3}$). \textit{Atmos} requires a trace amount of inert N$_{2}$ ($<$0.1\%) to prevent known numerical instability; but, this does not affect the observed atmospheres or dominant photochemical processes that appear in our modeling suites.} Our chemical reaction network (listed on the GitHub repository) was created by adopting the H$_{2}$O, O$_{2}$, CO$_{2}$, and SO$_{2}$ reaction networks presented by \citet{Lincowski2018trappist} for TRAPPIST-1 and Venus models \citep[used also by][for TRAPPIST-1c]{Lincowski2023t1catms}.}
\peerreview{In \textit{Atmos}, surface boundary conditions for a given species can be set as a} constant mixing ratio, a constant outgassing flux, or a surface deposition velocity. Top of atmosphere boundary conditions may be given as a constant escape flux or an effusion velocity; for Hydrogen specifically, diffusion-limited escape rates are calculated automatically. 

In this work, we set boundary conditions as constant fluxes or deposition/effusion velocities based on literature constraints and allow the model to determine the resulting surface pressure that arises from the balance of sources and sinks in a given simulation. As we are interested in determining if thin atmospheres can be sustained by outgassing alone, a novel approach is used is to treat the surface pressure as a dependent variable, rather than fixing it in a given atmosphere. \peerreview{Note, unlike all other gases considered in this work, we do consider a fixed trace amount of SO$_{2}$ in certain experiments. However, this mixing ratio never exceeds 1\% and does not significantly constrain the equilibration of the atmospheric pressure under the influence of outgassing, escape, and photochemistry.}

\subsection{Radiative Transfer Modeling SMART and LBLABC} \label{subsec:SMARTMethods}

We generate spectra with the Spectral Mapping Atmospheric Radiative Transfer (\textit{SMART}) model, a 1D line-by-line, multistream, fully multiple scattering radiative transfer model \citep[][]{Meadows1996smart,Crisp1997smart,Robinson2017smart} that has been validated against the observed spectra of several solar system bodies including Earth \citep[][]{Robinson2011epoxiearth}, Mars \citep[][]{Tinetti2005smartmars}, Titan \citep[][]{Robinson2014smarttitan}, and Venus \citep[][]{Arney2014smartvenus,Lincowski2021phosphine}. \textit{SMART} adopts the monochromatic multiple scattering solver, DISORT \citep[][]{Stamnes1988disort,Stamnes2000disort}. Furthermore, \textit{SMART} has been used to model the preindustrial and Archean Earth \citep[e.g.,][]{Arney2016pods,Currie2023moretolife} and it has been used extensively to model a variety of atmospheres in the TRAPPIST-1 system \citep[][]{Lustig2019detectT1,Gialluca2021characterizing,Gialluca2023corrigendum,Lincowski2018trappist,Lincowski2023t1catms,Zieba2023t1cjwst,Meadows2023t1biosigs}. Within a specified wavelength range, \textit{SMART} can generate planetary spectra at a specified resolution given a pressure-temperature profile, gas mixing ratios, UV-visible and collisionally induced absorption cross sections. Visible and infrared vibration-rotation absorption coefficients for gases generated by the Line-By-Line ABsorption Coefficients (\textit{LBLABC}) model \citep[][]{Meadows2018ProximaB} using the HITRAN line database \citep[][]{Gordon2022hitran2020}, a stellar spectrum, and wavelength-dependent surface albedo. In this work we include Rayleigh scattering, but neglect clouds and hazes. 

\subsection{The VPL Climate Model} \label{subsec:ClimateMethods}


The \textit{VPL Climate} Model is a terrestrial radiative-convective equilibrium climate model that can either be used in 1D or 1.5D (two column day-night sides) \citep[][]{Robinson2018life,Lincowski2018trappist,Lincowski2023t1catms}. This climate model is based on the previously described SMART radiative transfer model \citep[][]{Meadows1996smart}. \textit{VPL Climate} uses time stepping to advance the temperature profile of an atmosphere towards equilibrium, defined by the 1D surface-atmosphere thermodynamic energy equation. Heating rates are determined via the Linearized Flux Evolution method \citep[LiFE,][]{Robinson2018life}, which uses spectrally resolved Jacobians generated by SMART to calculate changes in radiative fluxes in response to changes in temperature. Whenever atmospheric properties evolve outside of the linear range of Jacobians, they are recalculated by SMART. Furthermore, mixing length theory is used to calculate convective heat fluxes and heating rates (treatment fully described in \citealp{Lincowski2018trappist}, with updates by \citealp{Meadows2023t1biosigs}).

The 1.5D, or two column, mode of \textit{VPL Climate} is used to generate profiles for the day and night sides of a terrestrial planet; the methods of this mode are described by \citet{Lincowski2023t1catms}. The 1.5D mode of \textit{VPL Climate} additionally calculates day-night heat transport through horizontal winds. This 1D version of \textit{VPL Climate} has been validated against 3D GCMs \citep[][]{Meadows2018ProximaB} and more critically, Earth and Venus profiles \citep[][]{Lincowski2018trappist,Meadows2023t1biosigs}, and the two-column mode has been validated against GCM-generated phase curves \citep[][]{Lincowski2018trappist,Lincowski2023t1catms}. Both the 1D and two-column version of \textit{VPL Climate} have also been used specifically to model planets in the TRAPPIST-1 system \citep[e.g.,][]{Lincowski2023t1catms,Meadows2023t1biosigs}. In this work, the two-column mode is used only for the interior planets of the TRAPPIST-1 system (b, c, and d) which may have dayside emission data in hand or are hot enough for it to be feasibly collected in the future. 

\subsection{\peerreview{Input Volatile Outgassing Rates}} \label{subsec:OutgassingModel}

\peerreview{Although present data on the TRAPPIST-1 planets does not provide direct constraints on geologic properties or outgassing, the volatile outgassing rates used as input in this study are taken from \citet{Thomas2025outgassrates}, who place theoretical boundaries on these quantities consistent with observations of the system \citep[e.g.,][]{Agol2021ttv} and knowledge of solar system bodies. While the densities of the planets are known to within 5\% accuracy \citep[][]{Agol2021ttv}, the space of possible interior compositions and structures consistent with these observed densities is highly degenerate \citep[e.g.,][]{Agol2021ttv,Elkins2008corelessplanet}. Other key information related to volatile outgassing is similarly unconstrained, such as the magma emplacement rate and bulk mantle volatile abundances. Given these uncertainties, the true interior structures, and related outgassing rates, for the TRAPPIST-1 planets are unknown.}

\peerreview{\citet{Thomas2025outgassrates} placed theoretical bounds on the TRAPPIST-1 outgassing rates by statistically exploring a broad range of interior scenarios spanning the full range of conditions that are theoretically possible on terrestrial planets and empirically observed in the solar system (on Earth, Mars, Venus, and Io). First, the model samples the mantle water content from negligible values (0.01 wt\%) up to the maximum value a terrestrial planet can sustain without becoming thermodynamically unstable \citep[$\sim$1 wt\%, according to][]{Guimond2023mantleminerology}. Next, the amount of water partitioned into the magma is calculated, limited by either the water abundance or solubility, both of which are calculated under the range of physical conditions observed during the production of Earth’s mantle-derived magmas \citep[][]{Sigurdsson2015volcanoes}. The magma emplacement rate, which is the most uncertain, is either sampled from thermal evolution models of TRAPPIST-1 \citep[0.01 -- 8500 km$^{3}$yr$^{-1}$;][]{Krissansen2022predictionsT1Interior} which span from negligible to an order of magnitude higher than Io, or it is fixed at values measured for Earth, Mars, and Io. In the present study, however, we restrict our analysis to samples that adopt magma emplacement rates derived from the TRAPPIST-1 thermal evolution models. Equilibrium degassing from the emplaced magma is calculated under the range of chemical conditions observed in the solar system \citep[e.g., magma oxygen fugacity ranging defined by FMQ-4 to FMQ+4, magma temperature ranging from 873 K to 1973 K, etc;][]{Thomas2025outgassrates}. Finally, additional filters are applied to incorporate empirical information on the TRAPPIST-1 system, such as strong observational constraints that preclude large H$_{2}$-dominated atmospheres \citep[][]{deWit2018trappist1Observe,Wakeford2019trappist1g} implying that H$_{2}$ cannot be the most abundant outgassed species.}

\peerreview{Given the broad phase space and uncertainties considered, the \citet{Thomas2025outgassrates} outgassing rates vary widely. For example, when all parameters are sampled over their full ranges, the water outgassing rate on T-1c can span $\sim$6 orders of magnitude. Therefore, we sample from these distributions in our modeling approach by initially testing the 2.5$^{th}$, 50$^{th}$, and 97.5$^{th}$ percentile values (median and lower/upper 2$\sigma$) of the \citet{Thomas2025outgassrates} outgassing rate distributions for all 7 planets.} 

\subsection{Model Framework} \label{subsec:ModelFrameworkMethods}

We develop a python wrapper which takes a novel approach to coupling \textit{Atmos}, \textit{VPL climate}, \textit{LBLABC}, and \textit{SMART} by automating a single atmosphere model run to convergence and parallelizing multiple models within an input phase space of atmospheric source and sink rates. 
To identify atmospheres which are the equilibrium state between outgassing and vigorous escape on these planets, a novel routine is added to update the surface pressure of a given atmosphere simulation searching for convergence between the photochemical and climate models. \peerreview{During one modeling pass through \textit{Atmos}, the pressure and number density of each model layer is fixed. Because of this, although mass is conserved in the atmosphere, the total mixing ratio of a layer (sum of mixing ratios for all species) may depart from 1. For example, if a layer begins with a number density of 1 cm$^{-3}$ and only contains 1 O$_{2}$ molecule, the total mixing ratio, equal to the mixing ratio of O$_{2}$, would be 1. If this O$_{2}$ molecule was then photolyzed to 2 O atoms, the number density of the layer would remain fixed and therefore the total mixing ratio, now equal to the mixing ratio of atomic O, would be 2. In other words,} 
if the cumulative atmospheric sources provide more or insufficient material needed for the \peerreview{set} pressure, the total mixing ratio of all gases may indicate this by being greater than or less than 1. In many past studies focused on matching best-fit models to observations \citep[e.g.,][]{Lincowski2023t1catms}, boundary conditions have been set by fixed mixing ratios, which allow the model to then infer the required outgassing or loss flux to maintain a total mixing ratio near 1 for a chosen pressure. This study improves on past work by moving away from fixed mixing ratio boundary conditions to allow the surface pressure of the atmosphere to equilibrate based on input fluxes that have been identified in the literature as realistic for the TRAPPIST-1 planets (see \S \ref{subsec:InputSpaceMethods}). We leverage this new functionality to determine whether thin atmospheres, which are still consistent with JWST observations \citep[][]{Greene2023t1b,Zieba2023t1cjwst,ih2023constrainingb,Lincowski2023t1catms,lim2023atmospheric,Ducrot2025combined,Radica2025t1c}, are also theoretically plausible given realistic outgassing. 

\peerreview{As noted above, \textit{Atmos} adopts a fixed pressure or number density versus altitude profile during each single model iteration. 
To automatically adjust the surface pressure of a given atmosphere between iterations of \textit{Atmos}, we calculate a corrected number density profile by scaling the fixed number density profile reported by the previous model run by the total mixing ratio (sum of all species mixing ratios), which acts as a renormalization factor. This corrected number density profile is integrated over altitude to find a column density, and subsequently a column mass, from which a revised surface pressure is derived. The pipeline will iterate between running \textit{Atmos} and updating its surface pressure until the surface pressure is changing by $<$3\%; note that model convergence is further described \S\ref{subsubsec:SingleModelConvergenceCriteriaMethods}.}

\peerreview{When adjusting the surface pressure,} we first find the corrected total number density per atmospheric layer from the reported number density and volume mixing ratios (which may be greater or less than 1):
\begin{equation} \label{Eq:nzCorr}
    n_{corr,z} = n_{report,z} \ \sum_{i} X_{i,z} \ .
\end{equation}

\peerreview{Here, for a given altitude layer $z$, the number density reported by the previous model run ($n_{report,z}$) is scaled by the total mixing ratio of that layer (the sum of all $i$ species mixing ratios, $X_{i,z}$) --- this provides the corrected number density for that layer, $n_{corr,z}$.} 
We then find the column number density by integrating the corrected number density \peerreview{profile} over altitude:
\begin{equation}
    N_{col} = \int_{z}^{\infty} n_{corr,z} \ dz \ .
\end{equation}
\peerreview{Note, we complete integration in the model using \textit{Numpy}'s trapezoidal integration scheme \citep[][]{Harris2020numpy}.} The column number density can be converted to the column mass using the atmospheric mean molecular weight, $\mu$, and Avogadro's number, $N_{Avo}$:
\begin{equation}
    \sigma_{col} = \frac{N_{col} \ \mu \ 10^{-3}}{N_{Avo}} \ ,
\end{equation}
assuming $\mu$ is in the units of g/mol and $N_{col}$ is in the units of m$^{-2}$, the factor of $10^{-3}$ converts the column mass to units of kg/m$^{2}$. Finally, the corrected surface pressure can be found with the column mass and the planetary gravity, $g$:
\begin{equation} \label{Eq:PsurfCorr}
    P_{surf,corr} = \sigma_{col} \ g \ .
\end{equation}
After creating an atmosphere model with \textit{Atmos}, our framework will calculate the corrected surface pressure (Eq. \ref{Eq:PsurfCorr}) and rerun \textit{Atmos}, if necessary (following the critera in \S \ref{subsubsec:SingleModelConvergenceCriteriaMethods}). 

\subsubsection{Phase Space Exploration Strategy} \label{subsubsec:ExploreStrategyMethods}


To organize our exploration of source-sink parameter space, we break up our approach into three phases. In the first phase, we focus on atmospheres that are fueled only by water outgassing, and are depleted in carbon or sulfur bearing species. This phase begins with an initial simulation suite that is standardized for all planets (inputs described in \S\ref{subsec:InputSpaceMethods}), which includes H$_{2}$O outgassing, O$_{3}$ and H$_{2}$O$_{2}$ dry deposition, and H$_{2}$, H, O, and O$_{2}$ TOA loss. The initial simulation suite contains 64 simulations per planet, though many of these simulations may ultimately not be viable. Based on the results of the initial simulation suite, we vary specific sources and sinks to push the model towards stable, converged atmospheres.

The second phase of our exploration repeats this process considering a combination of both H$_{2}$O and CO$_{2}$ outgassing. In addition to the sinks previously described in the H$_{2}$O outgassing only phase, we also consider CO dry deposition and CO$_{2}$ TOA escape. At the end of both the first (H$_{2}$O outgassed) and second (H$_{2}$O-CO$_{2}$ outgassed) phase of exploration, we compile all stable atmospheres across all simulation suites and test the addition of trace SO$_{2}$. Trace SO$_{2}$ is added as a fixed mixing ratio of 100 ppm, 0.1\%, or 1\% at the surface, so the balance of SO$_{2}$ source to sink is automatically adjusted in the bottom layer to maintain the set trace mixing ratio. Above the surface layer, SO$_{2}$ is subject to the standard photochemical model, i.e., the mixing ratio is not fixed or constrained at any other layer. Every simulation that is computationally converged is publicly available on GitHub\footnote{https://github.com/mgialluca/T1Atmospheres\_Gialluca2026}, containing the pressure (or altitude) dependent mixing ratios of each species in the atmosphere, as well as any available spectra that were generated. 

\subsubsection{Single Atmosphere Model Convergence \peerreview{\& Coupling}} \label{subsubsec:SingleModelConvergenceCriteriaMethods}

We employ nested local convergence checks to determine global convergence across three stages: photochemical modeling, surface pressure adjustment, and climate modeling. 
The wrapper begins by iteratively running the \textit{Atmos} photochemical model, with an initial state and selected boundary conditions, until it reaches local photochemical convergence.
\peerreview{Local convergence of the photochemical model is dependent on both the final timestep size, and on the change in gas abundances across a run. Since \textit{Atmos} uses a variable timestepping method, the size of the timestep increases as the atmosphere approaches equilibrium and the rates of change in gas abundances decrease. Once the timestep has reached 10$^{17}$ seconds ($>$3 Gyr), indicating the atmosphere has achieved stability over long timescales, the first \textit{Atmos} convergence criterion is met. The other criterion we employ for iterated local convergence of \textit{Atmos} requires that the summed fractional changes in abundance for all absorbers present at $\geq$0.1\% remains less than 1, thereby testing the stability of primary constituents from the beginning to the end of the model run. 
When both the timestepping criterion (must be $>$10$^{17}$ seconds) and the cumulative fractional abundance change criterion are met, local \textit{Atmos} convergence is achieved.}

After \textit{Atmos} finds convergence, the surface pressure is adjusted, and \textit{Atmos} is run to local photochemical convergence again. This process of \peerreview{iterating between} the surface pressure adjustment and \textit{Atmos} is repeated until the surface pressure changes by less than 3\% \peerreview{following the convergence of \textit{Atmos}}. Once the surface pressure is stable, line-by-line absorption coefficients are created with \textit{LBLABC} and passed to the \textit{VPL Climate} model, which is iteratively run to local climate convergence. Local convergence of the climate model occurs when the heating rate in the troposphere is less than 0.09 K/day, and all atmospheric layers are flux balanced within 1 W/m$^{2}$. After local climate convergence is achieved, the resulting temperature profile is passed back to \textit{Atmos}, and the process restarts. 

\peerreview{When transitioning from the photochemical to the climate portion of the pipeline, \textit{VPL Climate} receives the gas abundance profiles for the most radiatively active species previously calculated by \textit{Atmos}. The climate model then updates the temperature profile and the eddy diffusion rates, which are passed back to the photochemical model for the next iteration of the pipeline. In summary, the photochemical model updates gas abundance profiles and the climate model updates temperature and eddy diffusion profiles; these quantities are exchanged between the two models as the pipeline iterates towards global convergence. This method of iterating photochemical and climate models separately is commonly used for modeling planetary atmospheres \citep[e.g.,][]{Arney2016pods,Lincowski2023t1catms,Agundez2025atmmodel}.}

\begin{figure}
    \centering
    \includegraphics[width=0.7\textwidth]{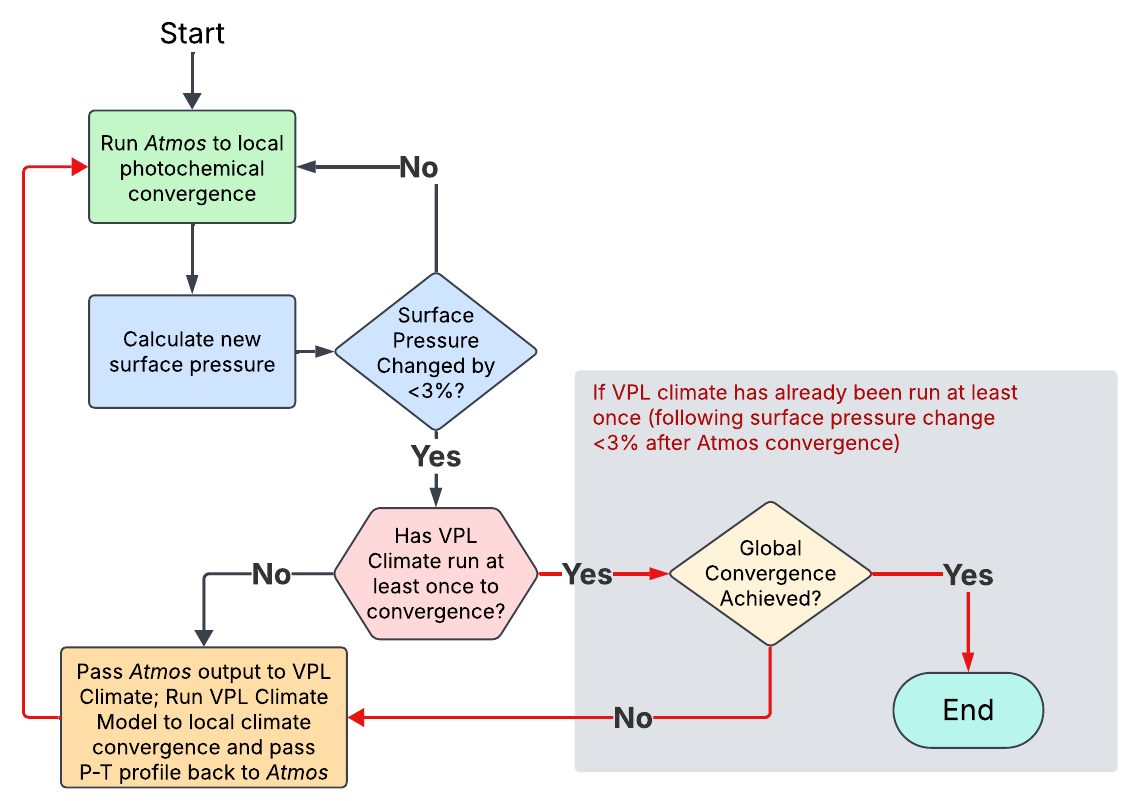}
    \caption{Flowchart describing how a single atmosphere model is run to convergence through the coupled photochemical-climate modeling. This is described in \S \ref{subsubsec:SingleModelConvergenceCriteriaMethods}.}
    \label{fig:ConvFlowChart}
\end{figure}

Global convergence checks occur after the surface pressure adjustment process once a full loop is completed (i.e., photochemical modeling through climate modeling with local convergence). Global convergence is achieved when, on a single pass through all modeling stages, the climate model finds local convergence on the first attempt, and the surface pressure never changes by more than 3\%. A flowchart illustrating this iterative photochemical-climate modeling process is shown in Figure \ref{fig:ConvFlowChart}.

Once global convergence is achieved in the coupled photochemical-climate portion of the modeling framework, transmission spectra are generated for the terminator portion of the planet (i.e., global average). In the case of planets that possess emission data (T-1b and c) or planets that may be reasonably accessible in the future with emission observations (T-1d), we further calculate day-night climate profiles and generate emission spectra. To produce the day and night side temperature profiles, the \textit{VPL Climate} model is used in 2 column mode and must find convergence where 
all atmospheric layers are flux balanced within 1 W/m$^{2}$ for \textit{both} the day and night side profiles. If 2-column convergence cannot be found, the atmosphere is rejected from being a possible solution. With the day and night profiles, transmission and emission spectra are then generated by SMART.

\subsubsection{Input Priors} \label{subsec:InputSpaceMethods}

We aim to generously encompass the phase space of plausible boundary conditions by defining broad ranges of source and sink flux rates informed by the literature. In this work, we focus on secondary atmospheres generated by water and CO$_{2}$ outgassing, with and without trace SO$_{2}$, based on the current best-fitting atmospheric cases from JWST observations \citep[][]{Greene2023t1b,lim2023atmospheric,ih2023constrainingb,Zieba2023t1cjwst,Lincowski2023t1catms,Radica2025t1c,Piaulet2025t1d}. The primary source for our atmospheres is either water outgassing alone, or mixed H$_{2}$O-CO$_{2}$ outgassing. All outgassing rates are constrained by \citet{Thomas2025outgassrates}, who used a magmatic outgassing model informed by solar system terrestrial bodies to define the region of outgassing rates that may be plausible for the TRAPPIST-1 planets. We allow outgassing rates to be sampled between the 2$\sigma$ lower and upper values of the probability distribution given by \citet{Thomas2025outgassrates} for a given planet.

There are 4 different sinks used in all atmosphere models: ozone (O$_{3}$) and hydrogen peroxide (H$_{2}$O$_{2}$) dry deposition at the surface, and atomic (O) and molecular oxygen (O$_{2}$) escape at the top of the atmosphere (TOA). Atmospheres that possess CO$_{2}$ outgassing also include CO$_{2}$ TOA escape and CO dry deposition. These loss processes were selected as common and strong sinks found on Earth and Mars \citep[e.g.,][]{Zahnle2008marsdepositonmodel,Constant2008coco2Quebec,Stein2014wintertimeCO,Seinfeld2016atmbook,Clifton2020ozonedepositionreview,Gronoff2020atmospheric} relative to other loss process related to water and its daughter photochemical products. Additionally, diffusion-limited H and H$_{2}$ escape are automatically included in \textit{Atmos}, as is the case with past work using this photochemical model \citep[e.g.,][]{Meadows2018ProximaB,Lincowski2018trappist,Lincowski2023t1catms,Meadows2023t1biosigs}. When defining deposition or effusion velocities, the following equation relates this velocity to total flux (cm$^{-2}$ s$^{-1}$) for a species ($s$):
\begin{equation} \label{eq:velocitytoflux}
    F_{s} = v \ n_s \ ,
\end{equation}
where $v$ is replaced by the deposition ($v_{dep}$) or effusion ($v_{eff}$) velocity (cm/s), and $n_s$ is the number density (cm$^{-3}$) of that species at the bottom or top layer of the atmosphere for surface deposition or TOA escape, respectively. For any given simulation, the deposition and effusion velocities of applicable sink processes are fixed, and the exact resulting loss flux will vary with the species' number densities (i.e., as composition and/or surface pressure change). 

On Earth, deposition velocities of O$_{3}$ and H$_{2}$O$_{2}$ vary based on surface type (e.g., ocean, desert, forest, etc.) and time of day. While continental O$_{3}$ deposition averages around 0.4 cm/s \citep[][]{Seinfeld2016atmbook}, this may be an overestimate for surfaces low in organic matter, such as the irradiated soils of the TRAPPIST-1 planets. For instance, the Sahara Desert shows values closer to 0.1 cm/s \citep[][]{Gusten1996ozonesahara,Clifton2020ozonedepositionreview}. Snow-covered regions also exhibit lower O$_{3}$ deposition, with velocities typically between 0 -- 0.1 cm/s range \citep[][]{Seinfeld2016atmbook,Clifton2020ozonedepositionreview}. Over the Northeastern US, ozone deposition velocities were found to range from 0.7 -- 0.8 cm/s at midday to 0.1 -- 0.2 cm/s at night, while H$_{2}$O$_{2}$ deposition velocities ranged from 1.6 -- 2 cm/s at midday to 0.6 -- 0.9 cm/s at night \citep[][]{Walcek1987theoretical}. Average H$_{2}$O$_{2}$ deposition velocities are about 0.5 cm/s over continents, 1 cm/s over oceans, and 0.32 cm/s over snow  \citep[][]{Seinfeld2016atmbook}. \citet{Zahnle2008marsdepositonmodel} models the past and present atmospheres of Mars using a 1D photochemical model, they found that the best fitting depositional velocities for ozone and H$_{2}$O$_{2}$ tend to be much lower than Earth, ranging from 0.01 to 0.03 cm/s for both gases. To account for this broad range of possibilities, our initial simulation suites begin by testing deposition rates of 0.4 and 0.02 cm/s for both O$_{3}$ and H$_{2}$O$_{2}$.

Carbon monoxide (CO) dry deposition on Earth is responsible for around 10 -- 15\% of the global atmospheric CO loss \citep{Stein2014wintertimeCO}. Depending on the surface composition, deposition velocities of CO range from negligible up to $\sim$0.1 cm/s for modern Earth \citep[][]{Constant2008coco2Quebec,Stein2014wintertimeCO}. However, as CO deposition is strongly biologically mediated on Earth, we assume that CO deposition will be less efficient in the abiotic environments considered in this study. \secondpeerreview{Since it is unknown empirically how CO deposition may operate on the TRAPPIST-1 planets, we test both CO depositional velocities of 0, and marginal velocities up to 0.01 cm/s.}


The TOA escape boundary conditions of atomic O, O$_{2}$, and CO$_{2}$ (when applicable) are set by effusion velocities to allow the exact escape flux to respond to changing pressure and species concentrations during a simulation. This is necessary because \peerreview{specifying a fixed} escape flux will prevent a simulation's sink processes from responding to a changing atmosphere, which often leads to numerical instability. However, setting an effusion velocity can lead to widely ranging final escape fluxes due to the dependence on exact atmospheric composition, pressure, and mixing ratio profiles\peerreview{, which affects species' number densities and ultimately the escape flux via Equation \ref{eq:velocitytoflux}}. In our results, we present the net escape fluxes of atmospheres found using this effusion velocity approach. In our discussion, we assess whether this inferred flux is reasonable in comparison to past studies.

\peerreview{A given effusion velocity may lead to widely varying escape fluxes, depending on final atmospheric composition. To provide some physical grounding for our ranges of effusion velocities, we consider some comparisons to Venus and Mars here. If we take the number density of atomic oxygen to be $\sim$10$^{10}$ -- 10$^{11}$ cm$^{-3}$ at 120 km ($\sim$10$^{-7}$ bars) in Venus' atmosphere \citep[][]{Brecht2012VenusatomicO,Limaye2017thermalvenus}, the effusion velocity would have to be 10$^{-4}$ -- 10$^{-5}$ cm/s to achieve an escape velocity of $\sim$10$^{24}$ -- 10$^{25}$ s$^{-1}$, similar to what is seen for neutral atomic oxygen sputtering and photochemical escape on Venus and Mars \citep[][]{Gronoff2020atmospheric}. In another example, if the number density of atomic oxygen at the exobase of Mars is taken to be $\sim$10$^{6}$ -- 10$^{8}$ \citep[][]{Qin2020marsAtomicO}, the effusion velocity would have to be 0.001 -- 1 cm/s to achieve escape rates of $\sim$10$^{24}$ -- 10$^{25}$ s$^{-1}$.}

\peerreview{Given there is very limited data on the TRAPPIST-1 system, and that the heightened stellar activity may favor stronger escape processes \citep[e.g.,][]{Howard2023T1flaringJWST}, we begin by selecting effusion velocities near 0.01 -- 1 cm/s and ultimately test effusion velocities as low as 10$^{-8}$ cm/s. More specifically,} for interior planets (T-1b, c, and d), \peerreview{we begin the initial simulation suite by testing} effusion velocities of 0.01 and 1 cm/s for atomic oxygen, and 0.01 and 0.1 cm/s.
For planets inside of and exterior to the habitable zone (T-1e, f, g, and h), we adopt 0.01 and 0.1 cm/s for atomic oxygen, and 0.01 and 0.05 cm/s for molecular oxygen effusion velocity \peerreview{in the initial simulation suites}. When transitioning to atmospheres that also contain CO$_{2}$ outgassing, we initially test 0, 0.01, and 0.1 cm/s for CO$_{2}$ effusion velocities. As described previously, based on the results of the initial simulation suite, we allow sources and sink rates to vary as we target specific atmospheric pressures and stable atmospheric configurations. For example, if an atmosphere fails to be generated against strong loss processes, the effusion velocity may be slightly lowered in the subsequent simulation suite. \peerreview{For T-1e in particular, we test a much broader range of effusion velocities --- with a lower limit down to 10$^{-8}$ cm/s --- to test if this would make a significant difference, but we found that it did not provide significantly different outcomes than an effusion velocity of 10$^{-6}$ cm/s.}

Table \ref{tab:initialSinks} provides the \peerreview{range of surface and TOA sink strengths that are tested across all simulation suites in this study}.
For all stable and converged atmospheric models found in our H$_{2}$O and H$_{2}$O-CO$_{2}$ outgassing suites, we go on to explore SO$_{2}$ as a trace gas. For every stable atmosphere we add a fixed mixing ratio of SO$_{2}$ at the surface of 100 ppm, 0.1\%, and 1\%. These are then run through our photochemical-climate modeling pipeline, including the surface pressure re-adjustment routine. \peerreview{While SO$_{2}$ in these cases is set as a trace fixed mixing ratio, this does not significantly restrict the surface pressure, which still varies considerably due to the outgassing of bulk constituents and subsequent photochemical processes.}


\setlength{\tabcolsep}{10pt}
\begin{table}[]
    \centering
    \normalsize
    \begin{tabular}{|c||c|c|c||c|c|c|}
        \cline{2-7}
         \multicolumn{1}{c|}{}& \multicolumn{6}{|c|}{Range of Tested Atmospheric Sink Strengths}\\
        \cline{2-7}
        \multicolumn{1}{c|}{} & \multicolumn{3}{|c||}{Effusion Velocity [cm/s]} & \multicolumn{3}{|c|}{Deposition Velocity [cm/s]} \\
         \hline
        Planet & O & O$_{2}$ & CO$_{2}$ & O$_{3}$ & H$_{2}$O$_{2}$ & CO \\
        \hline
        \hline
        \hspace{-35pt}\makecell{Innermost\\Planets\\(T-1b \& c)} & 0.001 -- 1.0 & 0.001 -- 0.1 & 0 -- 0.1 & 0.02 -- 0.4 & 0.02 -- 0.4 & 0 -- 0.01 \\
        \hline
        T-1d & 0.001 -- 1.0 & 10$^{-6}$ -- 0.1 & 0 -- 0.1 & 0.02 -- 0.4 & 0.02 -- 0.4 & 0 -- 0.01 \\
        \hline
        T-1e & 10$^{-8}$ -- 0.1 & 10$^{-8}$ -- 0.05 & 0 -- 0.1 & 0.02 -- 0.4 & 0.02 -- 0.4 & 0 -- 0.01 \\
        \hline
        \hspace{-35pt}\makecell{Outermost\\Planets\\(T-1f, g, \& h)} & 0.001 -- 0.1 & 10$^{-6}$ -- 0.05 & 0 -- 0.1 & 0.02 -- 0.4 & 0.02 -- 0.4 & 0 -- 0.01 \\
        \hline
    \end{tabular}
    \caption{\peerreview{The strengths of TOA (effusion velocities, cm/s) and surface (deposition velocities, cm/s) atmospheric sinks tested across all modeling suites. The innermost planets (T-1b \& c) and the outermost planets (T-1f, g, \& h) used the same ranges, but the two planets straddling the habitable zone (T-1d \& e) were allowed to vary more widely (see text for more details).} The final escape fluxes calculated from this effusion velocity approach are presented in detail and discussed further in the paper; in general the resulting escape fluxes are found to be within a reasonable range when compared to solar system bodies and past work on the TRAPPIST-1 system \citep[e.g.,][]{Dong2018trappistescape}.}
    \label{tab:initialSinks}
\end{table}




\subsection{Fitting Models to Data} \label{subsec:goodnessoffitMethods}

In this work we place a particular focus on the extent to which modeled planetary atmospheres fit JWST data, when available. For T-1b, emission data at 15 and 12.8 $\mu$m \citep[][]{Greene2023t1b,Ducrot2025combined} is available, as well as transmission data from JWST/NIRISS across 2 visits \citep[][]{lim2023atmospheric}. For T-1c, emission data at 15 $\mu$m is currently available \citep[][]{Zieba2023t1cjwst}, as well as transmission data from JWST/NIRISS across 2 visits \citep[][]{Radica2025t1c}. For both T-1b and c, thermal phase curves further provide a 15$\mu$m nightside emission point, as well as another dayside 15$\mu$m measurement \citep[][]{Gillon2025phase}. For T-1d, transmission data from JWST/NIRSpec across 2 visits are currently available \citep[][]{Piaulet2025t1d}. Finally, for T-1e, transmission data from JWST/NIRSpec across 4 visits are currently available \citep[][]{Glidden2025t1e,Espinoza2025t1e}. Note, the more recent data of \citet{Allen2025t1ePrelimResults} was published after the results were completed in this work, while we use available data to analyze models conducted here, since these more recent T-1e observations would not change atmospheric interpretation or detection confidences, we stick to using the \citet{Espinoza2025t1e,Glidden2025t1e} data in our analyses.

When comparing an atmospheric model to available data, we quote a ``$\sigma$ confidence", which is by definition the number of standard deviations from the mean of a normal two-sided probability distribution that corresponds to the $p$-value found from a $\chi^{2}$ goodness-of-fit test. First we calculate the $\chi^{2}$ value:
\begin{equation}
    \chi^{2} = \sum_{i} \left( \dfrac{O_i - M_i}{\sigma_i} \right)^{2} \ ,
\end{equation}
where $O_i$ is the i$^{th}$ data-point and $M_i$ is the corresponding model point, and $\sigma_i$ is the uncertainty on $O_i$. With $\chi^{2}$ we can then calculate the $p$-value, which is the probability of obtaining a $\chi^{2}$ value greater than or equal to the calculated value assuming degrees of freedom equal to the number of observations ($\nu$ = $N_i$):
\begin{equation}
    p = \mathbb{P} \left( \chi^{2}_{\nu} \geq \chi^{2}_{\nu,obs} \right) \ .
\end{equation}
Finally we can use this $p$-value to arrive at a $\sigma$ confidence that the model matches the data.
In practice, we calculate these values using the Scipy.stats.chi2 package cumulative distribution function \citep[][]{Virtanen2021scipy}.

\subsubsection{Correlation Coefficient} \label{subsubsec:CorrelationCoeffMethods}

Given that standard 'goodness-of-fit' data-model comparisons neglect the correlation of adjacent wavelength bins, we employ the Pearson Correlation Coefficient as an additional metric to compare modeled spectra to data, as it may provide sensitivity to multiple, correlated, spectral features from a given molecular species through the slope of the spectra.
Traditionally when matching individual models to observation outside of a full retrieval, the $\chi^{2}$, $p$-value, and/or $\sigma$ fit methods described above are used. However, these methods treat each data-model point independently, when in reality spectral features are correlated; for example, if significant CO$_{2}$ is present in an atmosphere, one should see an absorption feature above the baseline at both 4.3 and 15$\mu$m. \peerreview{Thus, more information on how the model fits the data can be discerned from considering the general trend of the spectrum as a whole.} In other words, a good fitting model should be positively correlated to the data points (i.e., when the data points indicate a larger transmission depth, the model's transmission depth should also increase). 

To quantify the additional information that can be derived from analyzing how a modeled spectra correlates to the observations, we use the Pearson Correlation Coefficient (PCC). The PCC ranges from -1 to 1 with -1 being perfectly negative correlated, 0 being no correlation, and 1 being perfectly positively correlated. Ideally, a good fitting model should have a PCC approaching 1, indicating the model predicts absorption when there is potential absorption seen in the data compared to its baseline. We use the PCC to compare modeled transmission spectra to JWST data in this study (applicable for T-1b, c, d, and e). Since emission photometry only provides 1 data point, a correlation across a dataset cannot be established. 

For a single model to data comparison, the PCC is defined as:
\begin{equation} \label{eq:PCC}
    \rho_{CC} = \frac{N_i \left( \sum\limits_{i} O_i M_i \right) - \left( \sum\limits_{i} O_i \right) \left( \sum\limits_{i} M_i \right)}{\sqrt{ \left[ N_i \left( \sum\limits_{i} O_i^{2} \right) - \left( \sum\limits_{i} O_i \right)^{2} \right] \left[ N_i \left( \sum\limits_{i} M_i^{2} \right) - \left( \sum\limits_{i} M_i \right)^{2} \right] }} \ ,
\end{equation} 
where $O_i$ is the i$^{th}$ observation, $M_i$ is the corresponding model point, and $N_{i}$ is the total number of observations. Note, Equation \ref{eq:PCC} does not inherently account for when there are observational uncertainties ($\sigma_i$). To fold in the observational uncertainty for a given model-to-data comparison, we first take 500 randomly sampled observational datasets, taken assuming each data point has an asymmetrical normal distribution defined by that point's value and upper and lower uncertainty. Then, for each of the randomly sampled datasets, we calculate the PCC with the particular modeled spectrum being tested (calculating 500 PCCs), and we finally report the 50$^{th}$ percentile value of the set of PCCs as the answer. Since this metric has not been commonly used in this field, we provide a proof of concept for how this can maximize information in Appendix \ref{AppendixSec:PCCProof}.

\section{Results} \label{sec:Results}

Here we present the results of our atmosphere simulations, focusing on stable results that represent plausible atmospheres. We then simulate spectra from the stable atmospheres to compare to existing JWST observations of the TRAPPIST-1 planets. 

\subsection{Properties of the Stable Atmospheres} \label{subsec:photochemistryResults}

Our study tested grid sweeps across a broad parameter space of different boundary conditions, which led to several stable atmosphere types, and many non-viable combinations. Table \ref{tab:NumberOfAtms} shows the number of atmospheres that were stable and converged (from each outgassing source, and with or without SO$_{2}$), as well as the total number of tested atmospheres, for all planets. The \secondpeerreview{possible} failure modes for unstable atmospheres were 1) atmospheric runaway (where outgassing significantly exceeded loss processes), 2) complete atmospheric loss ($<$10$^{-10}$ bars, escape processes dominated), or 3) unstable surface pressure or temperature (oscillating between two values without converging). \secondpeerreview{In general, simulations would most commonly fail due to atmospheric collapse or numerical instability (oscillating physical parameters), and these two failure modes occurred at nearly equal frequency}. In the failure state of atmospheric runaway, the final pressure would typically be between 10 -- 500 bars, but the main indicator of this state is that pressure would increase by $\gtrsim$1 bar on every model iteration, indicating that gases are building up. \secondpeerreview{Notably, no atmospheres supported by water outgassing only (no present CO$_{2}$) were stable for T-f, g, or h, as the cold temperatures led to rapid collapse caused by freeze out.}

\peerreview{While we present a set of stable atmospheric solutions for these planets, since we did not use a statistical sampling method, we cannot quantify the overall likelihood of an atmosphere over an airless case, and we cannot conclude if one composition is statistically more likely over another.} Furthermore, as shown by Table \ref{tab:NumberOfAtms}, T-1f, in particular had a notably lower number of stable atmospheres than even the more outer planets in the system (T-1g and h). In the case of T-1f, a higher number of models failed to achieve `global convergence' as there would be more oscillations in a given pass through the \textit{VPL Climate} model. It may be that a more relaxed convergence method could be used in the future, allowing a higher number of atmospheres for T-1f, which we further elaborate on in the discussion. 

\setlength{\tabcolsep}{5pt}
\begin{table}[]
    \centering
    \normalsize
    \begin{tabular}{|c||c|c|c|c||c|c|}
    \hline
        Planet & \hspace{-30pt}\makecell{Total Stable\\without SO$_{2}$} & \hspace{-30pt}\makecell{H$_{2}$O\\Outgassed\\(without SO$_{2}$)} & \hspace{-30pt}\makecell{H$_{2}$O-CO$_{2}$\\Outgassed\\(without SO$_{2}$)} & \hspace{-30pt}\makecell{Total Tested\\(without SO$_{2}$)} & \hspace{-30pt}\makecell{Total Stable\\with SO$_{2}$} & \hspace{-30pt}\makecell{Total Tested\\with SO$_{2}$}\\
        \hline\hline
        b & 21 & 15 & 6 & 448 & 11 & 63 \\
        \hline
        c & 114 & 76 & 38& 448 & 97 & 342 \\
        \hline
        d & 175 & 134 & 41 & 448 & 163 & 525 \\
        \hline
        e & 224 & 200 & 24 & 703 & 606 & 672 \\
        \hline
        f & 4 & 0 & 4 & 512 & 0 & 12\\
        \hline
        g & 17 & 0 & 17 & 512 & 4 & 51 \\
        \hline
        h & 27 & 0 & 27 & 448 & 60 & 81 \\
        \hline
    \end{tabular}
    \caption{Number of total atmospheres tested and stable atmospheres for each planet as a function of outgassing composition (H$_{2}$O or H$_{2}$O-CO$_{2}$), and with or without the addition of a fixed mixing ratio of SO$_{2}$ at the surface (either 100 ppm, 0.1\%, or 1\%). While many stable atmospheres were found for T-1c, d, and e, the more outer planets (T-1f, g, and h) produced unstable results for water outgassing only, but several stable results when CO$_{2}$ was included. T-1b produced a sample of stable atmospheres, though less than T-1c, d, and e because of its proximity to the host star leading to a more irradiated environment. Since SO$_{2}$ was tested in 3 ways for every stable atmosphere, the total number of tested SO$_{2}$-bearing atmospheres is just 3$\times$ the second column, but this is explicitly calculated in the last column of the table. Including SO$_{2}$ also led to a number of stable atmospheres for all planets except T-1f, with a preference for smaller SO$_{2}$ amounts (i.e., 100 ppm versus 1\%) as this would lead to a smaller perturbation from an already stable state. T-1f likely did not produce stable SO$_{2}$-bearing atmospheres because of undersampling; there were only 4 atmospheres stable for T-1f originally, so only 12 configurations with SO$_{2}$ were tested.}
    \label{tab:NumberOfAtms}
\end{table}

\subsubsection{Atmospheric Composition}\label{subsubsec:compositionresults}

Across all of the stable atmospheres generated by outgassing that were identified, we found a pattern of 6 different atmospheric archetypes, though not every planet was found to have stable models for all 6. The first 3 archetypes could occur in atmospheres generated through water outgassing alone: steam-dominated ($>$90\% H$_{2}$O volume mixing ratio, VMR, at the surface); dry O$_{2}$-dominated sustained by H$_{2}$O outgassing ($>$90\% O$_{2}$ VMR at the surface); and mixed H$_{2}$O-O$_{2}$. The other 3 atmosphere archetypes occur as a consequence of combined H$_{2}$O-CO$_{2}$ outgassing: mixed CO$_{2}$-CO, dry O$_{2}$-dominated sustained by H$_{2}$O-CO$_{2}$ outgassing ($>$90\% O$_{2}$ VMR at the surface), and mixed O$_{2}$-H$_{2}$O-CO$_{2}$. Note, two separate archetypes are listed as O$_{2}$-dominated, but they have different trace gas compositions, as one is supported by H$_{2}$O outgassing alone, and the other by a mix of H$_{2}$O-CO$_{2}$ outgassing.  Furthermore, the `dominant' constituent in the mixed atmospheric types (H$_{2}$O-O$_{2}$, CO$_{2}$-CO, and O$_{2}$-H$_{2}$O-CO$_{2}$) may vary case-by-case. 
For the T-1b and T-1d simulations, stable solutions were found for all archetypes except for H$_{2}$O-CO$_{2}$ outgassed O$_{2}$-dominated. The T-1c simulations found stable solutions for all archetypes. The T-1e simulations provided stable solutions for everything except the H$_{2}$O-dominated and mixed H$_{2}$O-O$_{2}$ archetypes. None of the H$_{2}$O outgassed archetypes were stable in our simulations for T-1f, g, and h because the outgassed water would quickly freeze out due to the colder temperatures. However, all H$_{2}$O-CO$_{2}$ outgassed archetypes were stable for T-1g.   H$_{2}$O-CO$_{2}$ outgassed archetypes with the exception of mixed O$_{2}$-H$_{2}$O-CO$_{2}$ were stable for T-1f. Finally, for T-1h, only mixed CO$_{2}$-CO atmospheres were found to be stable.  

\begin{figure}
    \centering
    \includegraphics[width=\linewidth]{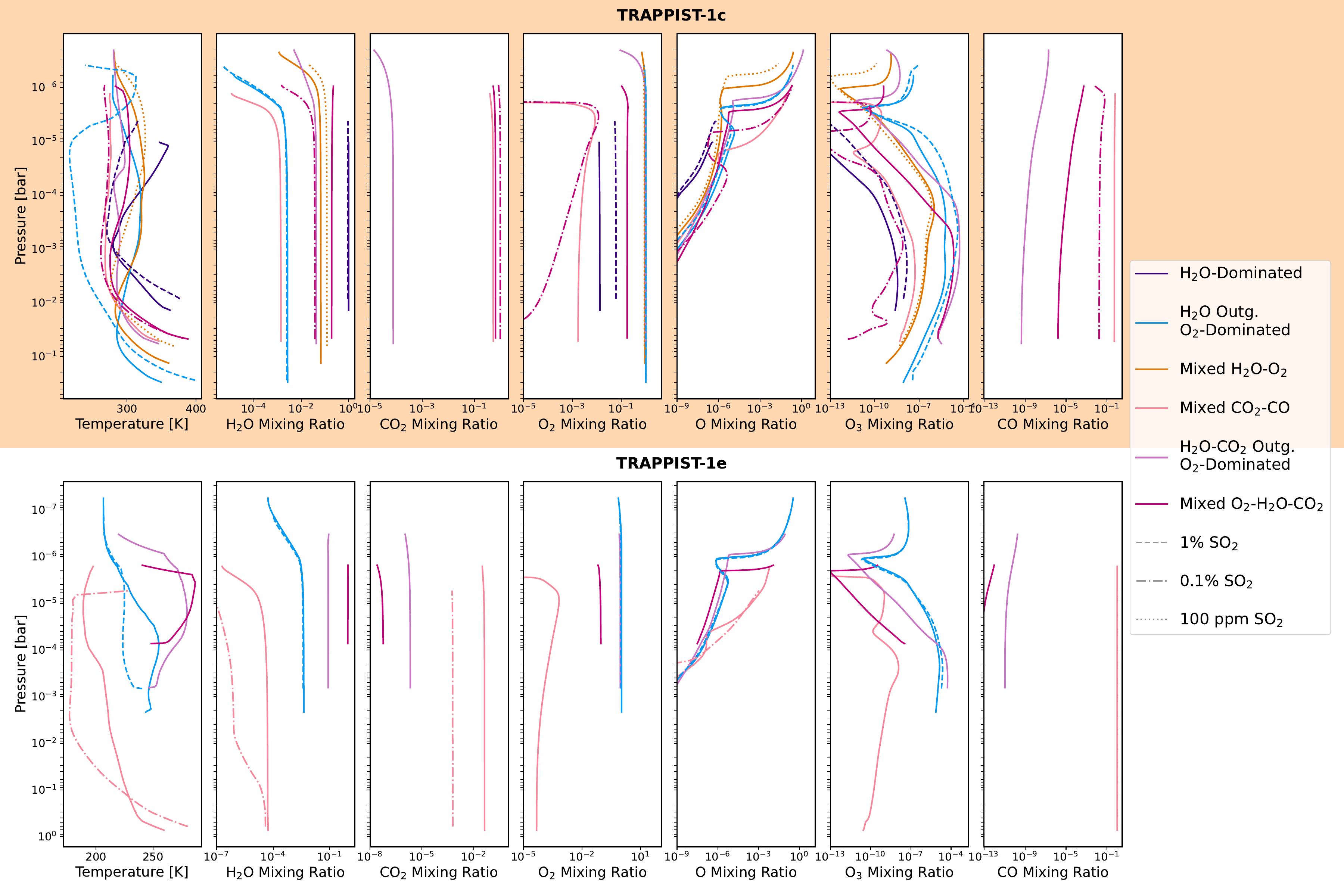}
    \caption{\peerreview{Examples illustrating the compositional and pressure-temperature range of the atmospheric archetypes found in this study. Temperature and gas mixing ratio profiles (as a function of atmospheric pressure) for H$_{2}$O, CO$_{2}$, O$_{2}$, O, O$_{3}$, and CO are shown for examples of every archetype atmosphere for T-1c and e.} The top row (peach colored background) shows profiles for T-1c and the bottom row (white colored background) shows T-1e profiles. When available, atmospheric cases are shown both without (solid lines) and with SO$_{2}$ contributions of either 1\% (dashed line), 0.1\% (dashed dotted line), or 100 ppm (dotted line) injected into the bottom layer.}
    \label{fig:CandEPhotochem}
\end{figure}

\begin{figure}
    \centering
    \includegraphics[width=0.9\linewidth]{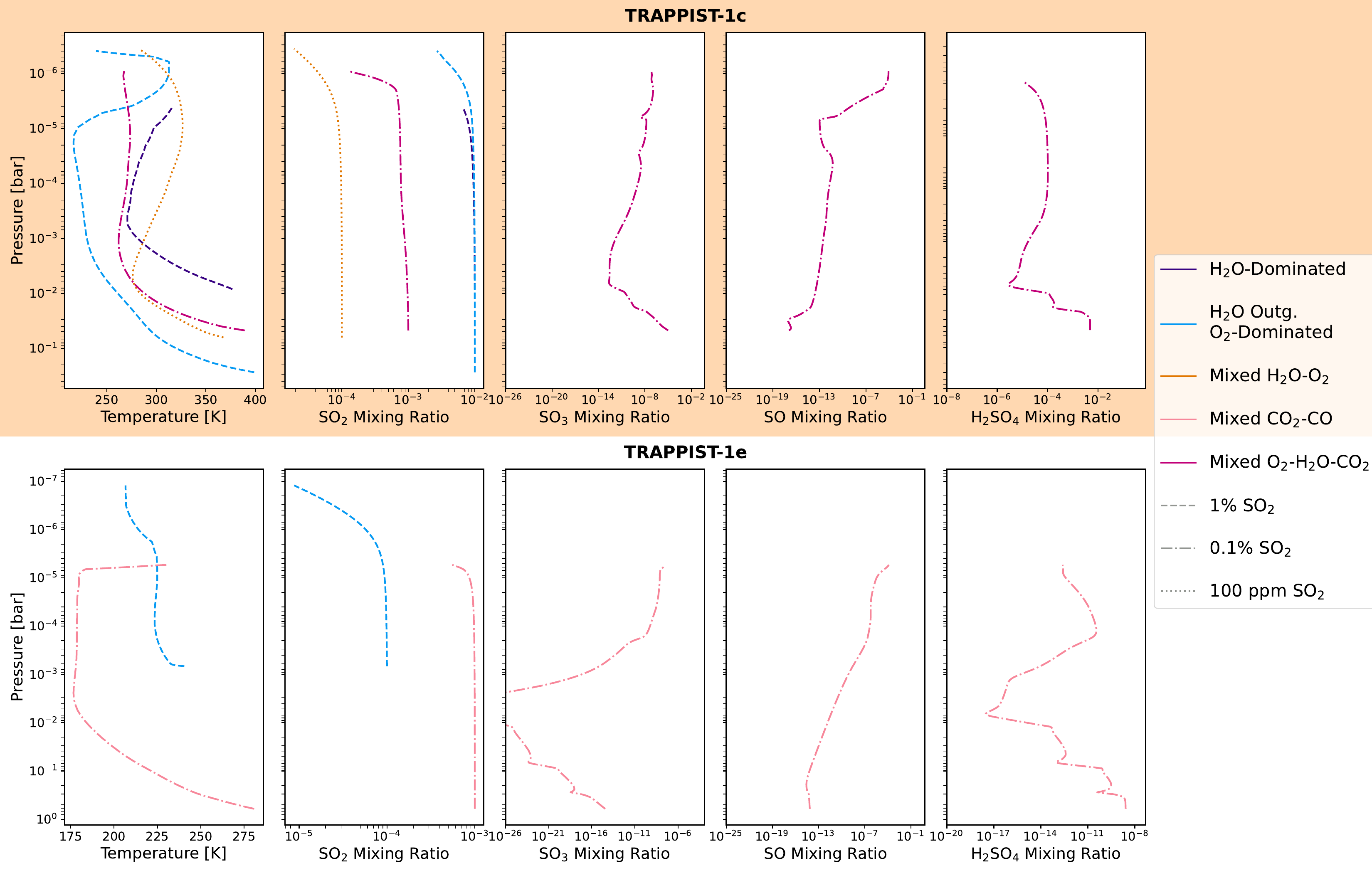}
    \caption{The volume mixing ratio profiles of SO$_{2}$ and daughter species for the same atmospheres given as examples in Figure \ref{fig:CandEPhotochem} for T-1c and e. Format is the same as Figure \ref{fig:CandEPhotochem}. \secondpeerreview{For T-1c and e, we show one example profile for every stable compositions for each of the two planets. Only the mixed CO-CO$_{2}$ and O$_{2}$-H$_{2}$O-CO$_{2}$ compositions had non-negligible abundances of SO$_{3}$, SO, and H$_{2}$SO$_{4}$, which is why no mixing ratio profiles are seen for these species in the other compositional cases.}}
    \label{fig:so2PTZPlot}
\end{figure}

Figure \ref{fig:CandEPhotochem} shows example cases of several key gas mixing ratio and temperature profiles as a function of atmospheric pressure for the atmospheric archetypes identified in this study. \peerreview{These plots show the diversity and range of outcomes from the coupled photochemical-climate model generated from the outgassing fluxes and modulated by the escape rates.} For each of TRAPPIST-1c and e (representative of the interior and outer planets), we show one case per stable atmospheric archetype, in addition to the same atmosphere with an SO$_{2}$ addition, if available. The largest stable addition of SO$_{2}$ to a given atmosphere is preferred, i.e., if adding 1\% SO$_{2}$ is unstable, but 0.1\% is stable, then the 0.1\% SO$_2$ result is shown. Additionally, for the SO$_{2}$-bearing examples, relevant sulfur species are shown in Figure \ref{fig:so2PTZPlot}. The abundance of SO$_{3}$, SO, and H$_{2}$SO$_{4}$ is negligible for all cases except the mixed O$_{2}$-H$_{2}$O-CO$_{2}$ atmosphere for T-1c and the mixed CO$_{2}$-CO atmosphere for T-1e. 

The temperature profiles for almost all of the T-1c atmospheres display an adiabatic drop-off in the lower atmosphere, followed by a near isothermal profile. The exception is for the H$_{2}$O-bearing atmospheres, with the H$_{2}$O-dominated atmosphere providing the most pronounced example, which show heating in the upper atmosphere that is likely due to water vapor absorption. The T-1e temperature profiles are more varied, with the mixed CO$_{2}$-CO atmosphere showing a similar behavior to the atmospheres of T-1c (drop-off and isothermal); the H$_{2}$O outgassed O$_{2}$-dominated showing heating in the lower atmosphere, followed by cooling and more isothermal behavior in the upper atmosphere;
and the H$_{2}$O-CO$_{2}$ outgassed O$_{2}$-dominated and the mixed O$_{2}$-H$_{2}$O-CO$_{2}$ atmospheres displaying heating in the lower atmosphere, followed by cooling in the mid to upper atmosphere. In general, the addition of SO$_{2}$ tends to lower the temperature of the atmosphere by reducing the water fraction. 

As the only two possible outgassed species, the mixing ratio of H$_{2}$O and CO$_{2}$, are mostly set by the source fluxes, particularly near the surface. The profiles of H$_{2}$O and CO$_{2}$ remain quite stable in every atmosphere, until the upper regions ($\sim$10$^{-6}$ bar) where photolysis leads to a steady drop-off in H$_{2}$O, and to a lesser degree, CO$_{2}$, and subsequent production of atomic O, ozone, and slightly increased CO. The water profile for the T-1e SO$_{2}$-bearing mixed CO$_{2}$-CO atmosphere (pink dash-dot line, second row of panels, in Figure \ref{fig:CandEPhotochem}) significantly deviates from all other cases, showing a steep depletion of H$_{2}$O in the lower atmosphere, likely from reactions with SO$_{3}$. 
The O$_{2}$ profiles across all atmospheres remain quite stable and constant, until the upper atmosphere where photolytic destruction and TOA escape causes a drop in O$_{2}$ and production of atomic O. The loss of O$_{2}$ in the upper atmosphere is especially stark for the mixed CO$_{2}$-CO atmospheres for both planets and the SO$_{2}$-bearing mixed O$_{2}$-H$_{2}$O-CO$_{2}$ atmosphere for T-1c. 
The atomic O and ozone profiles are strongly tied to the photolysis of water and O$_{2}$, and subsequent recombination processes (particularly for O$_{3}$). Given surface deposition is in the lowest atmospheric layer and ozone destruction via reactions with OH also dominate in the lower atmosphere, ozone increases in abundance from the surface, hitting a peak in the middle atmosphere ($\sim$10$^{-3}$ bars), and then decreasing towards the upper atmosphere. This slow decrease in ozone corresponds to an slow increase in atomic O, and both species show a sharp increase at the top of the atmosphere corresponding to the destruction of H$_{2}$O and O$_{2}$. Lastly, CO tends to increase very slowly \peerreview{as a function of altitude} when it is a non-negligible constituent in an atmosphere, this is likely due to depositional loss at the surface and heightened production through CO$_{2}$ destruction in the mid and upper atmosphere. \secondpeerreview{An exception to this behavior of CO is for the mixed CO-CO$_{2}$ atmospheres, which were often dominated by CO, and were only found to be stable for cases that had zero CO surface deposition; this is reflected in the relatively constant profile of CO as a function of altitude.}

\subsubsection{Pressure-Temperature Phase Diagrams} \label{subsec:PTDiagramsResults}

\begin{figure}
    \centering
    \includegraphics[width=\linewidth]{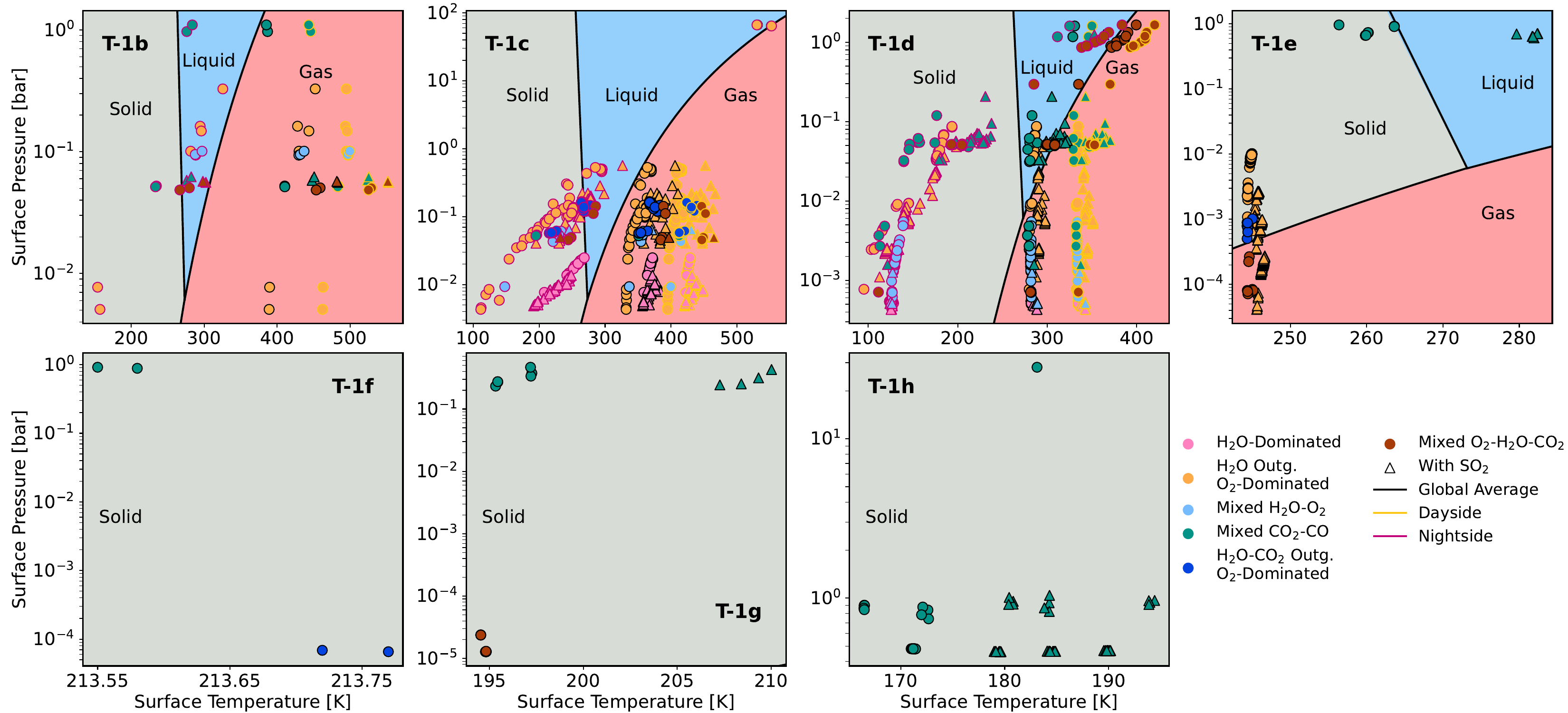}
    \caption{Surface pressure-temperature (P-T) points for each atmosphere in our work for each planet. Points are shown plotted over the phase diagram for water, where the grey, blue, and red shaded regions indicated solid, liquid, and gas phases. Points are colored by atmospheric composition (see figure legend). Globally averaged P-T pairs are shown with black borders, and dayside and nightside profiles for T-1b, c, and d are shown with yellow and magenta borders, respectively. Atmospheres containing SO$_{2}$ are shown by triangle points. T-1d and e both have atmospheres that may be fully contained in the liquid water region of the phase diagram.}
    \label{fig:PTDiagrams}
\end{figure}

Although potentially habitable surface temperatures are found for some of the atmospheric models we present, these atmospheres may also have extremely low pressures; in which case, the physical conditions required to support liquid water on the surface of a planet may not be met, despite the promising temperature.  To determine whether models with habitable pressure-temperature conditions exist, in Figure \ref{fig:PTDiagrams} we  plot the pressure-temperature pairs for all of our simulations, superimposed on the phase diagram for water. The color of points corresponds to the atmospheric composition, and the shape (circle versus triangle) denotes the lack, or addition, of SO$_{2}$. Globally averaged \peerreview{(black borders) }surface temperatures and pressures are shown for all planets, and \peerreview{day (yellow borders) and night (magenta borders) side} values are additionally shown for T-1b, c, and d. The water phase diagram is split into a grey, blue, and red region indicating solid, liquid, and gas phases of water. 

\peerreview{In terms of habitability, the globally averaged temperatures for T-1b and c are too hot and those for T-1f, g, and h are too cold, but several profiles of T-1d and e indicate potentially habitable pressure-temperature profiles. As expected, T-1b and c show very hot daysides and global averages, and very cold nightsides, many with hemispherically averaged temperatures below the freezing point of water. Some of the two interior planets' nightsides may even be temperate (between 273.15 -- 300 K), though in all of these cases we use hemispheric averages. However, these averages lack the spatial resolution to accurately assess the potential for spatially-localized regions of extreme cold that could induce atmospheric collapse of the nightside. All atmospheres for T-1f, g, and h are within in the coldest region of the phase diagram, indicating water is likely in ice form, which is consistent with our result that none of these planets could sustain an atmosphere from water outgassing alone.} 


\peerreview{T-1d and e exhibit a range of stable atmospheres that produce above freezing surface temperatures in the global average. T-1 d is a very interesting case, existing right on the inner edge of the potential habitable zone \citep[e.g.,][]{Gillon2017seven,Wolf2017climate3dtrappist,Meadows2023t1biosigs}; and} notably, a few $\sim$1 bar mixed CO$_{2}$-CO atmospheres for T-1d have nightside, global average, and dayside temperatures all within the liquid water region of the phase diagram. In the case of T-1e, we found that only mixed CO$_{2}$-CO atmospheres fall within the liquid water region, and that typically some addition of trace SO$_{2}$ was necessary to push the atmosphere solidly into this region.

\subsubsection{Atmospheric Pressure \& Escape Fluxes} \label{subsec:SurfPressEscapeFluxResults}



Nearly all stable atmospheres found in this study have thin ($<$1 bar) pressures and high O escape rates ($\sim$10$^{26}$ -- 10$^{30}$ s$^{-1}$). \peerreview{As shown in Figure \ref{fig:PTDiagrams},} the possible surface pressures for T-1b, c, and d range from 0.005 -- 1 bars, 0.005 -- 0.5 bars, and 0.0004 -- 2 bars, respectively, with the exception of two high pressure outlier models for T-1c discussed below. For T-1e, f, and g, the surface pressures range from approximately 10$^{-5}$ -- 1 bars, and for T-1h pressures range from 0.4 -- 1 bars with the exception of one high pressure outlier simulation discussed below. However, we caution that the outer planets (T-1f through h), as well as T-1b, may be under-sampled, as stable atmospheres were difficult to maintain for the particular input combinations and convergence restrictions that we tested (see Table \ref{tab:NumberOfAtms}).

There are a few surface pressure outliers for T-1c and h that likely possessed unrealistically low escape rates, allowing persistence of a massive atmosphere. For T-1c, there were two stable high-pressure ($\sim$60 bar) H$_{2}$O outgassed O$_{2}$-dominated atmospheres, which possessed O and O$_{2}$ escape rates approximately 1 order of magnitude less than many of the other, low pressure, O$_{2}$-dominated atmospheres. Although O$_{2}$ does not exhibit strong spectral features at JWST-accessible wavelength ranges for atmospheres of 1 bar or less \citep[][]{Meadows2023t1biosigs}, for higher pressure ($>3$ bars) O$_{2}$-dominated atmospheres, collisionally induced absorption from O$_2$ molecules can produce strong and broad features near 1.06 and 1.27 $\mu$m in transmission \citep{Misra2014using}. Since these O$_2$-O$_2$ features may be accessible to JWST for high enough pressures \citep{Lustig2019detectT1}, and the outlying 60-bar T-1c atmospheres provide $\sim$3$\sigma$ fits to the available data, these atmospheres are  still included in further analyses. For T-1h, there is one outlying high pressure, $\sim$30 bar, atmosphere which is nearly pure CO (produced by H$_{2}$O-CO$_{2}$ outgassing); this particular case was tested with negligible CO deposition and negligible CO$_{2}$ top-of-atmosphere (TOA) escape, so it was likely caused by an unrealistic build-up. However, since the CO deposition on exoplanets is extremely difficult to constrain, and because there is no public JWST data for T-1h, we decided to leave this unlikely high-pressure atmosphere in the remaining analyses of this study.  

\begin{figure}
    \centering
    \includegraphics[width=0.8\textwidth]{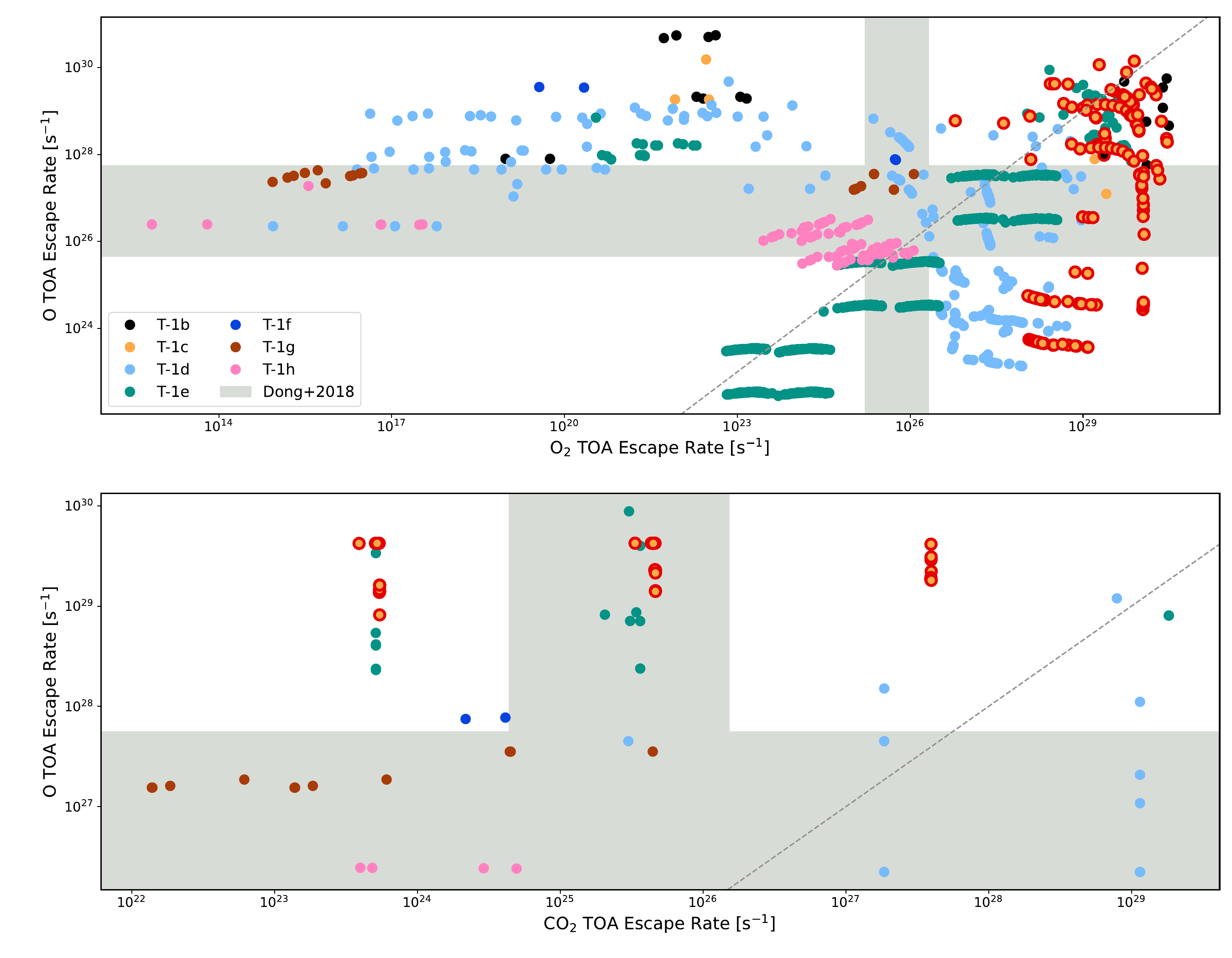}
    \caption{The escape fluxes (in s$^{-1}$) found across all stable atmospheres for all planets. Top panel shows the O versus O$_{2}$ escape flux, and the bottom panel shows the O versus CO$_{2}$ escape flux (for atmospheres including CO$_{2}$ outgassing). For T-1b and c, models that fit the available emission data to $<$3$\sigma$ are denoted by red rings. The grey regions indicated the range of escape rates (across all 7 planets) found by \citet{Dong2018trappistescape}. The models that best fit the T-1b and c emission data tend to have higher O and O$_{2}$ escape rates.}
  \label{fig:EscapeRates}
\end{figure}

The maximum total escape flux of O, O$_{2}$, and CO$_{2}$ across all atmospheres was $\sim$10$^{30}$ s$^{-1}$.
As previously described, we set TOA escape rates using an effusion velocity, as opposed to a strict fixed flux, to allow loss fluxes to respond to changing concentrations in the upper atmosphere; this is important because a fixed outgassing flux paired with fixed escape processes will not be able to dynamically evolve with the atmosphere (e.g., as composition or pressure changes). Here, we explore the resulting escape fluxes from our use of effusion velocities. Figure \ref{fig:EscapeRates} shows the equilibrium escape fluxes found for every stable atmosphere; the top panel shows the O versus O$_{2}$ escape flux and the bottom panel shows O versus CO$_{2}$. The dashed grey line on both panels indicates where the O and O$_{2}$ (top), or O and CO$_{2}$ (bottom), escape rates would be equal. Points with red rings highlight T-b and c modelled escape rates that produce atmospheres with spectra that are consistent with  available emission data to within $<$1$\sigma$ \citep[][]{Greene2023t1b,Zieba2023t1cjwst,Ducrot2025combined,Gillon2025phase}. For comparison, we show the range of ion escape rates for TRAPPIST-1 planets assuming a Venus-like atmosphere, as constrained by \citet{Dong2018trappistescape}, as grey regions in Figure \ref{fig:EscapeRates}. In general, the O and O$_{2}$ escape rates of our  best fitting T-1b and c models are high ($\sim$10$^{27}$ -- 10$^{30}$ s$^{-1}$), and are equal to, or up to 3 orders of magnitude greater than the maximum escape rates reported by \citet{Dong2018trappistescape} for a 1 bar Venus-like atmosphere. However, we note that neutral escape may be significantly larger (Brain \& Hinton, \textit{priv. comm.}) than the ion escape processes found by \citet{Dong2018trappistescape} and these high escape rates in our study were balanced by the high end of plausible outgassing rates calculated in \citet{Thomas2025outgassrates}. While we did not statistically sample the outgassing rate distributions at high resolution, in general, stable atmospheres in this study required outgassing rates above the 50$^{th}$ percentile value of the distributions from \citet{Thomas2025outgassrates}. 
Finally, in regards to the escape rates presented in Figure \ref{fig:EscapeRates}, we would like to note that the 'clumping' sometimes seen for groups of simulations is likely an artifact of the sampling method (grid sweeps), not evidence that certain escape rates are ultimately preferred.

\subsection{Atmospheric Spectra and Comparison with JWST Observations} \label{sec:AtmOverviewResults} 

\begin{figure}
    \centering
    \includegraphics[width=\linewidth]{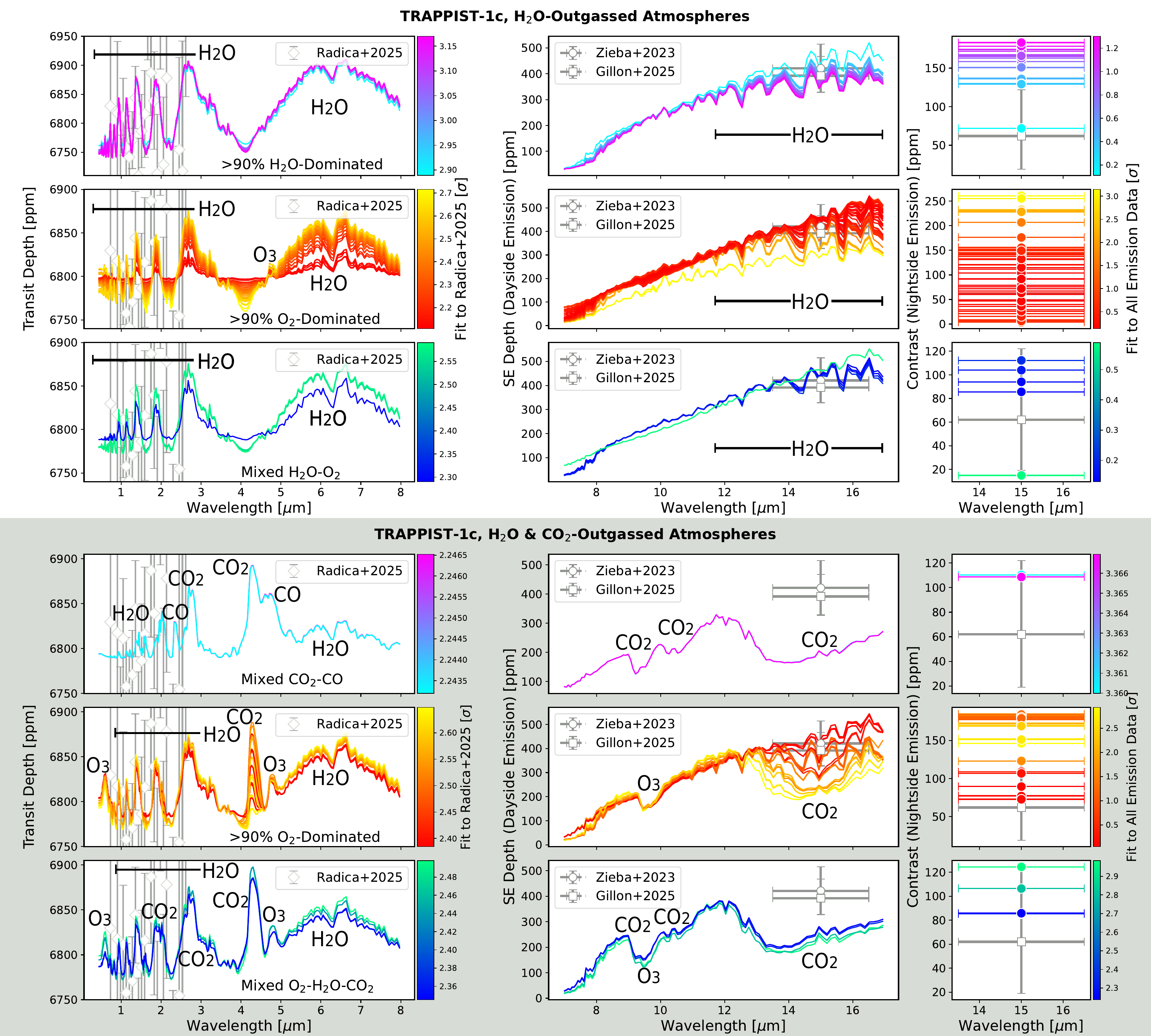}
    \caption{Transmission and emission spectra for all stable atmospheres found for T-1c (without trace SO$_{2}$). Top 3 rows (white background) and bottom 3 rows (grey background) show atmospheres sourced from H$_{2}$O and H$_{2}$O-CO$_{2}$ outgassing, respectively. Each row corresponds to a different atmospheric archetype (described in text), from top to bottom: H$_{2}$O-dominated, O$_{2}$-dominated (produced via H$_{2}$O outgassing), mixed H$_{2}$O$_{2}$, mixed CO$_{2}$-CO, O$_{2}$-dominated (produced via H$_{2}$O-CO$_{2}$ outgassing), and mixed O$_{2}$-H$_{2}$O-CO${2}$. Left column shows visible to near-infrared transmission spectra (0.001 -- 6.5$\mu$m), middle column shows secondary eclipse emission spectra at thermal wavelengths (8 -- 17$\mu$m), and right column shows nightside thermal emission at 15$\mu$m. Transmission spectra are colored by their fit ($\sigma$-deviation) to available T-1c transmission data \citep[][]{Radica2025t1c}, and emission data are colored by their combined fit to all available emission data \citep[15$\mu$m day and night points;][]{Zieba2023t1cjwst,Gillon2025phase}. Spectral features are labeled with the absorbing gas. The overall best fitting atmospheres for T-1c include lower pressure ($<$0.2 bar) O$_{2}$-dominated atmospheres (regardless of outgassing source) and mixed O$_{2}$-H$_{2}$O atmospheres.}
    \label{fig:T1cSpectraSummary}
\end{figure}

For each planet, we generated a summary figure that shows all the available model spectra, for comparison with known observational constraints from secondary eclipse, thermal phase curve and transmission spectroscopy measurements. In each figure, the spectra are colored by their fit to the available observational data, when applicable. We first discuss Figure \ref{fig:T1cSpectraSummary} where we show the spectral results for T-1c, 
which has stable atmospheres for all compositional archetypes described above. In Figure \ref{fig:T1cSpectraSummary}, the top three rows (white background color) show the H$_{2}$O outgassed archetypes while the bottom three rows (grey background color) show the H$_{2}$O-CO$_{2}$ outgassed archetypes. The first column of panels shows transmission spectra in the wavelength range of 0.001 -- 6.5 $\mu$m, the second column of panels shows secondary eclipse (day-side emission) spectra in the wavelength range of 8 -- 17 $\mu$m, and the third column of panels (for the T-1b and c spectral plots only) show night-side emission averaged over  the MIRI 15$\mu$m photometric bandpass, for comparison with available thermal phase curve measurements of the nightside emission \citep{Gillon2025phase}. 
The modeled transmission spectra are colored by their fit to the transmission data only \citep[for T-1c, this is from][]{Radica2025t1c}, while the emission spectra are colored by their joint fit to all available emission data, which can include secondary eclipse data and thermal phase curve data \citep[for T-1c,][]{Zieba2023t1cjwst,Gillon2025phase}. Spectral features in Figure \ref{fig:T1cSpectraSummary} (and subsequent spectra figures) are labeled by the absorbing gas.

The broad range of compositions for these atmospheres produce many characteristic spectral features, including for molecules that are chemically and photochemically generated from the outgassing products.  As can be seen in Fig. \ref{fig:T1cSpectraSummary}, for the water-dominated outgassing atmospheres their spectra are, perhaps predictably, dominated by water vapor features in both the near- and mid-infrared.  For the O$_2$-dominated atmosphere produced by photolysis of outgassed water, which has the lowest fraction of water vapor of the three water outgassing atmospheres, relatively weak features from photochemically-generated O$_3$ are also seen.  In contrast the atmospheres generated from H$_2$O/CO$_2$ outgassing show more complex spectra, because of the addition of the CO$_2$ and its photochemical byproducts such as CO and O$_3$. For these atmospheres the spectra show weaker features from H$_2$O, when compared to the H$_2$O outgassing cases, additional features from CO and strong CO$_2$ absorption, as well as stronger features from O$_3$, including at mid-infrared wavelengths, when compared to those seen in the H$_2$O-only outgassing cases.   

Considering the data-model comparison results for T-1c, the overall best-fit atmospheric archetypes for the available observations appear to be O$_{2}$-dominated regardless of outgassing source, and mixed H$_{2}$O-O$_{2}$ atmospheres (second, third, and fifth rows of Figure \ref{fig:T1cSpectraSummary}). Most atmospheres found here for T-1c are consistent with  transmission data \citep[][]{Radica2025t1c} to between 2 -- 3$\sigma$ but, given the extreme difficulty in perfectly removing stellar contamination from transmission observations of TRAPPIST-1 \citep[e.g.][]{Rackham2018TLSE,Zhang2018nirtransmissiontrappist,lim2023atmospheric,Rathcke2025stellarcorrect}, we conclude that there is not sufficient confidence to rule any of these atmospheres out. However, H$_{2}$O-dominated atmospheres (top row Figure \ref{fig:T1cSpectraSummary}) are slightly worse fits to the transmission data (at $\sim$3.1$\sigma$) than all other archetypes. Looking at comparisons with the emission data for T-1c \citep[][]{Zieba2023t1cjwst,Gillon2025phase}, which is more robust against stellar contamination, mixed CO$_{2}$-CO atmospheres (fourth row, second column, Figure \ref{fig:T1cSpectraSummary}) provide the worst fits at $>$3$\sigma$, followed by mixed O$_{2}$-H$_{2}$O-CO$_{2}$ atmospheres (bottom row, second column Figure \ref{fig:T1cSpectraSummary}) at 2 -- 3$\sigma$. Both of these atmosphere types produce strong absorption at 15$\mu$m from CO$_2$, and this is not seen in the data.  Although they were least preferred in transmission, H$_{2}$O-dominated atmospheres (top row, Figure \ref{fig:T1cSpectraSummary}) fit T-1c emission data moderately well from 0.1 -- 1.2$\sigma$, with the lower pressure atmospheres providing a better fit. The best-fitting atmospheres overall for T-1c are O$_{2}$-dominated atmospheres that had a surface pressure $<$0.2 bars
(second and fifth rows, Figure \ref{fig:T1cSpectraSummary}), and the mixed H$_{2}$O-O$_{2}$ atmospheres (third row, Figure \ref{fig:T1cSpectraSummary}), which all fit the emission data to $<$1$\sigma$.

\begin{figure}
    \centering
    \includegraphics[width=\linewidth]{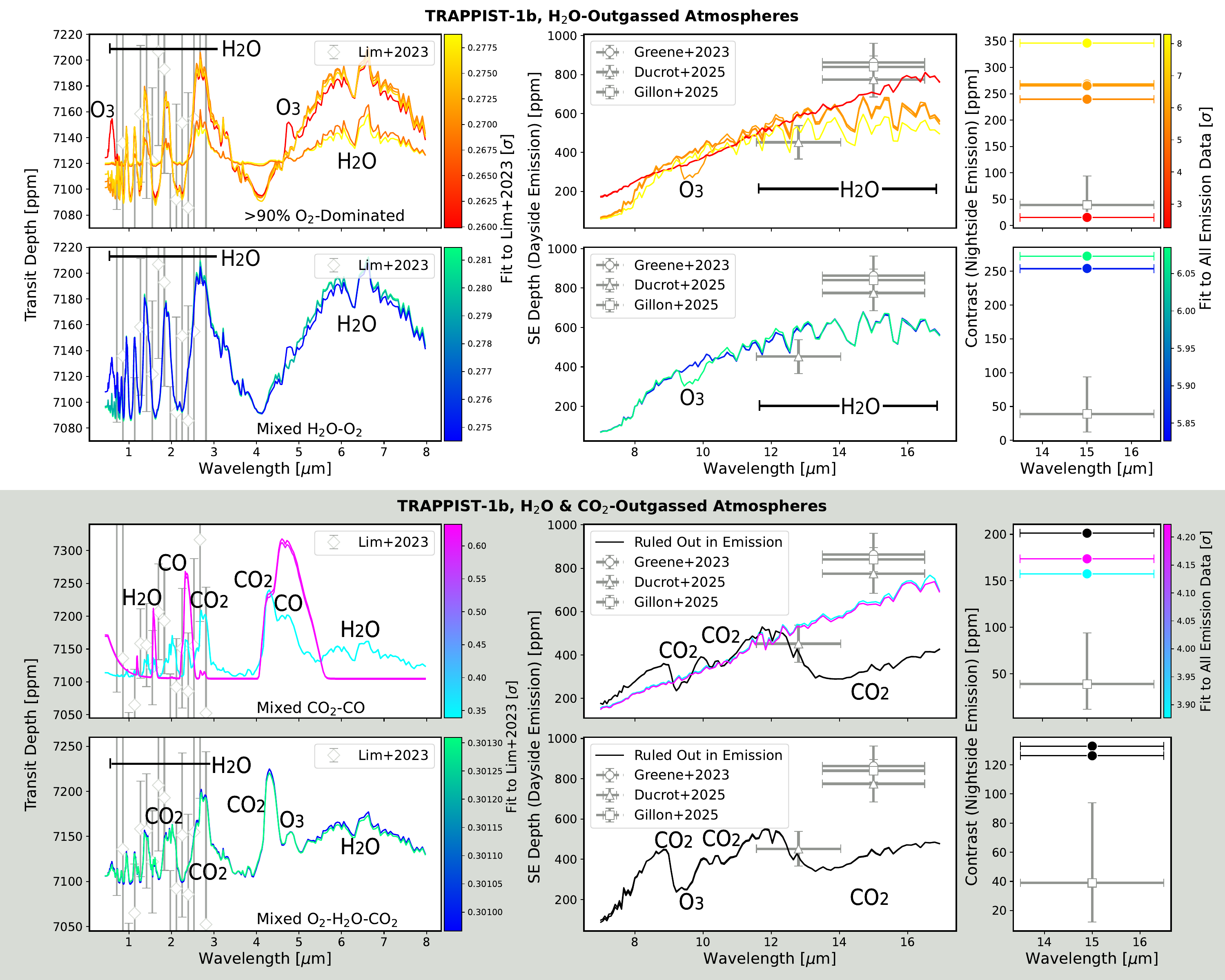}
    \caption{Transmission and emission spectra for stable atmospheres for T-1 b. transmission/emission columns and atmosphere archetypes (rows) are as for Figure \ref{fig:T1cSpectraSummary}, except that T-1b had no stable, realistic atmospheres of the H$_{2}$O-dominated or H$_{2}$O-CO$_{2}$ outgassed O$_{2}$-dominated archetypes, so those rows are not present. The only atmospheres that fit the T-1b emission data \citep[][]{Greene2023t1b,Gillon2025phase,Ducrot2025combined} to the $\sim$2$\sigma$ level are H$_{2}$O outgassed O$_{2}$-dominated atmospheres with pressures of $\sim$0.005 bars. The T-1b mixed CO$_{2}$-CO and mixed O$_{2}$-H$_{2}$O-CO$_{2}$ spectra shown in black indicate that they are confidently ruled out by emission data (fitting at a $\chi^2$ of $\sim$100).}
    \label{fig:T1bSpectraSummary}
\end{figure}

\begin{figure}
    \centering
    \includegraphics[width=\linewidth]{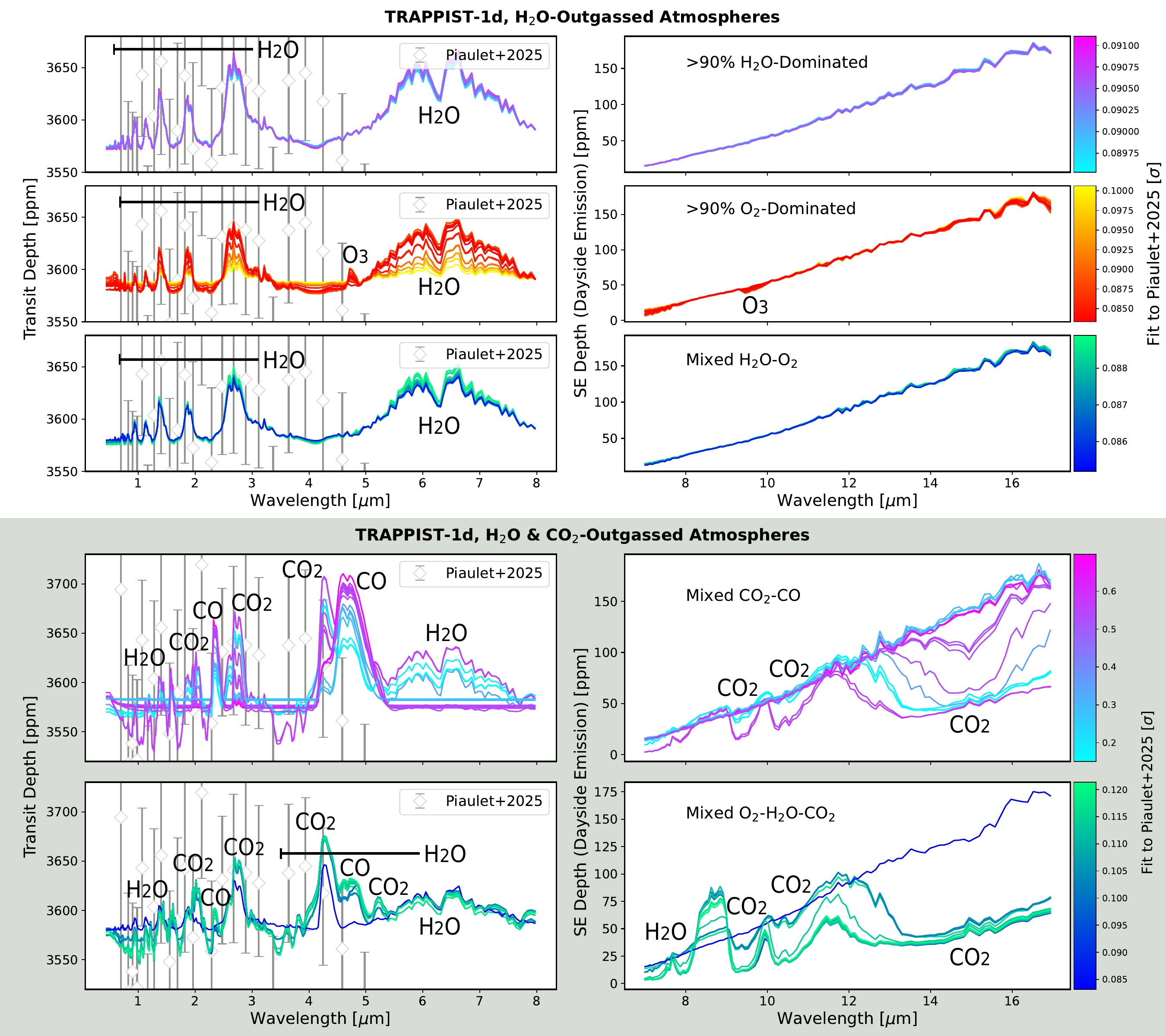}
    \caption{Same as Figure \ref{fig:T1cSpectraSummary} but for T-1d. However, T-1d had no stable H$_{2}$O-CO$_{2}$ outgassed O$_{2}$-dominated atmospheres, so that row is absent from the figure. Moreover, as T-1d has no published emission data, both transmission and emission spectra are colored by that model's fit to the available transmission data \citep[][]{Piaulet2025t1d}. All modeled atmospheres in this study fit the T-1d transmission data to $<$1$\sigma$.}
    \label{fig:T1dSpectraSummary}
\end{figure}

\begin{figure}
    \centering
    \includegraphics[width=\linewidth]{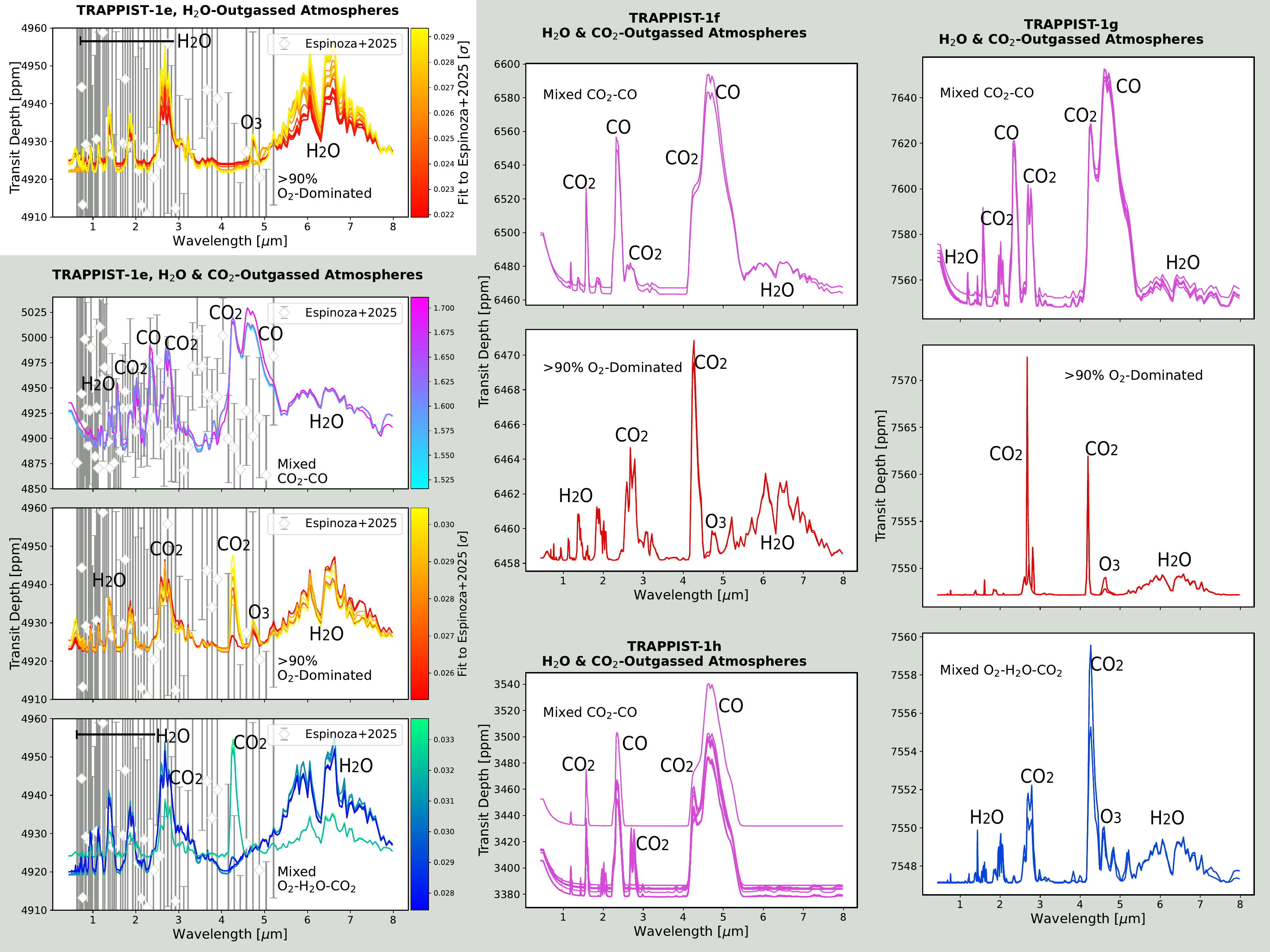}
    \caption{Similar to Figure \ref{fig:T1cSpectraSummary}, but contains the spectra for T-1e, f, g, and h. The left column shows transmission spectra for T-1e, and is laid out the same as the left column of Figure \ref{fig:T1cSpectraSummary}; spectra are colored by their fit to the recent T-1e transmission data \citep[][]{Espinoza2025t1e,Glidden2025t1e}. T-1e has no stable H$_{2}$O-dominated or mixed H$_{2}$O-O$_{2}$ atmospheres. The middle column shows transmission spectra for T-1g, the top two panels of the right column show transmission spectra for T-1f, and the bottom panel of the right column shows transmission spectra for T-1h. T1-f, g, and h are too cold to sustain any stable atmospheres supported by H$_{2}$O outgassing alone. T-1g has all H$_{2}$O-CO$_{2}$ outgassed archetypes, but T-1f is missing mixed O$_{2}$-H$_{2}$O-CO$_{2}$ atmospheres and T-1h only has stable CO$_{2}$-CO atmospheres.}
    \label{fig:T1efghSpectaSummary}
\end{figure}

The spectra summary figure for T-1b is given by Figure \ref{fig:T1bSpectraSummary}, the layout is the same as Figure \ref{fig:T1cSpectraSummary} except T-1b had no stable atmospheres for the H$_{2}$O-dominated and H$_{2}$O-CO$_{2}$ outgassed O$_{2}$-dominated archetypes, so those rows are not included. All stable atmospheres identified for T-1b match the available transmission data \citep[][]{lim2023atmospheric} to $<$1$\sigma$. However, the data error bars are \~60-100ppm, which is larger than the extent of many spectra features in this wavelength range, and large uncertainties in transmission may still be exacerbated by unremoved stellar contamination. In terms of emission data, almost every atmosphere archetype we modeled is ruled out to high ($>$4$\sigma$) confidence, except for two H$_{2}$O outgassed O$_{2}$-dominated atmospheres with pressures of $\sim$0.005 bars which fit the emission data at 2.3$\sigma$. 

The spectra summary figure for T-1d is given by Figure \ref{fig:T1dSpectraSummary}, the layout is the same as Figure \ref{fig:T1cSpectraSummary}, except T-1d has no published emission data at the time of this study, so all spectra in both the transmission and emission plots are colored by their fit to the available transmission data \citep[][]{Piaulet2025t1d}; T-1d also had no stable H$_{2}$O-CO$_{2}$ outgassed O$_{2}$-dominated atmospheres, so that row is absent from the figure. All atmospheres found for T-1d in this study fit the available transmission data \citep[][]{Piaulet2025t1d} to $<$1$\sigma$, but as above, the extent of the data error bars exceeded the size of spectral features in all cases. 

For T-1d, there was a very strong difference in the thermal features seen for atmospheres generated by the two different outgassing modes.  While the water outgassing atmospheres produced largely featureless thermal spectra, those stable atmospheres generated from CO$_2$-H$_2$O outgassing showed a broad range of features due to CO$_2$ absorption. Figure \ref{fig:T1dSpectraSummary} shows that subsets of the mixed CO$_{2}$-CO and mixed O$_{2}$-H$_{2}$O-CO$_{2}$ atmospheres have weaker absorption, and so noticeably higher emission near the 15$\mu$m band. For the mixed CO$_{2}$-CO atmospheres, the strongest to weakest absorption features occurred when  the surface mixing ratio of CO$_{2}$ ranged from $\gtrsim$1\% to $\lesssim$10 ppm. For the mixed O$_{2}$-H$_{2}$O-CO$_{2}$ atmospheres, most of the stable results had high ($>$90\%) CO$_{2}$ mixing ratios, leading to more absorption and low thermal emission at 15$\mu$m ($\sim$50 ppm). However, cases with only $\sim$20\% CO$_{2}$ and larger O$_{2}$ and H$_{2}$O contributions, the CO$_2$ absorption was weaker, and emission at 15$\mu$m was much higher. 

The modeled spectra for T-1e, f, g, and h are all shown on Figure \ref{fig:T1efghSpectaSummary}. The left column of Figure \ref{fig:T1efghSpectaSummary} shows atmospheres for T-1e, the top two panels of the middle column show atmospheres for T-1f, the bottom panel of the middle column shows atmospheres for T-1h, and the right-most column shows atmospheres for T-1g. We do not include emission spectra for these outer planets (T-1e through h) as their cold temperatures reduce the feasibility of emission photometry. The layout is the similar to the left-most column of Figure \ref{fig:T1cSpectraSummary}, for T-1e this includes coloring by the model fit to the available JWST transmission data \citep[][]{Espinoza2025t1e,Glidden2025t1e}. For T-1f, g, and h, where JWST data were not yet available at the time of this paper, we plot spectra in one color, based on the atmospheric archetype shown. 

Beginning with T-1e, no stable H$_{2}$O-dominated or mixed H$_{2}$O-O$_{2}$ atmospheres were found, but all other archetypes produced stable results. Furthermore, all modeled atmospheres fit the available transmission data \citep[][]{Espinoza2025t1e,Glidden2025t1e} to $<$1$\sigma$, except for the mixed CO$_{2}$-CO compositions, which still fit the data to $<$2$\sigma$. For T-1f, g, and h, no atmospheres sustained by H$_{2}$O outgassing, alone, were found to be stable, as their lack of other greenhouse gases and corresponding cold radiative-equilibrium temperatures lead to water condensation before substantial build-up or photolysis could occur. For T-1f, only stable mixed CO$_{2}$-CO and H$_{2}$O-CO$_{2}$ outgassed O$_{2}$-dominated atmospheres are found. For T-1g, all archetypes for H$_{2}$O-CO$_{2}$ outgassing are present; and for T-1h, only stable mixed CO$_{2}$-CO atmospheres were found. 

\subsubsection{Spectra with Trace SO$_{2}$} \label{subsubsec:TraceSO2OverviewResults}

For every stable atmosphere found in the H$_{2}$O only and H$_{2}$O-CO$_{2}$ outgassing suites, we tested 3 additional cases with the addition of fixed SO$_{2}$ mixing ratio of 100 ppm, 0.1\%, and 1\% at the bottom of the atmosphere; these SO$_{2}$-bearing models were then rerun through our photochemical-climate modeling pipeline. Following the strategy of the previous section (\S \ref{sec:AtmOverviewResults}), we provide a spectra summary figure that matches the layout of Figure \ref{fig:T1cSpectraSummary} and shows all available SO$_{2}$-bearing spectra for each planet. 
Notably, for planets with emission spectra, the addition of SO$_{2}$ creates a clear emission feature at $\sim$7.3$\mu$m and $\sim$8.7$\mu$m, increases emission seen just shortward of 15$\mu$m, and reduces the emission at longer wavelengths.

\begin{figure}
    \centering
    \includegraphics[width=\linewidth]{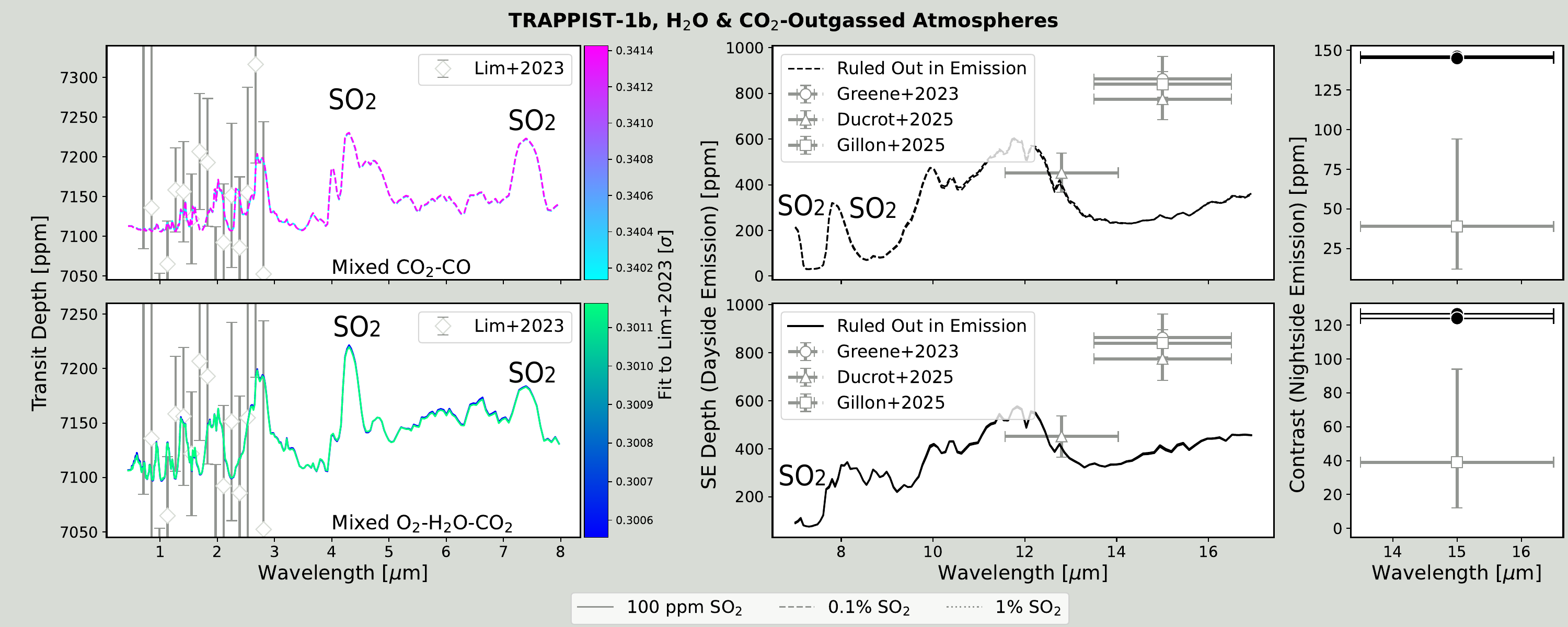}
    \caption{Same as Figure \ref{fig:T1cSpectraSummary} but for the SO$_{2}$-bearing atmospheres of T-1b. Solid, dashed, and dotted lines indicate that 100 ppm, 0.1\%, or 1\% SO$_{2}$, respectively, was added to the bottom layer of the atmosphere. When including SO$_{2}$, T-1b only had stable, realistic atmospheres for H$_{2}$O-CO$_{2}$ outgassed mixed CO$_{2}$-CO and mixed O$_{2}$-H$_{2}$O-CO$_{2}$ compositions. However, excessive heat redistribution rules out these compositions for T-1b based on the currently available emission data \citep[][]{Greene2023t1b,Ducrot2025combined,Gillon2025phase}.}
    \label{fig:SO2T1bSpectraSummary}
\end{figure}

\begin{figure}
    \centering
    \includegraphics[width=\linewidth]{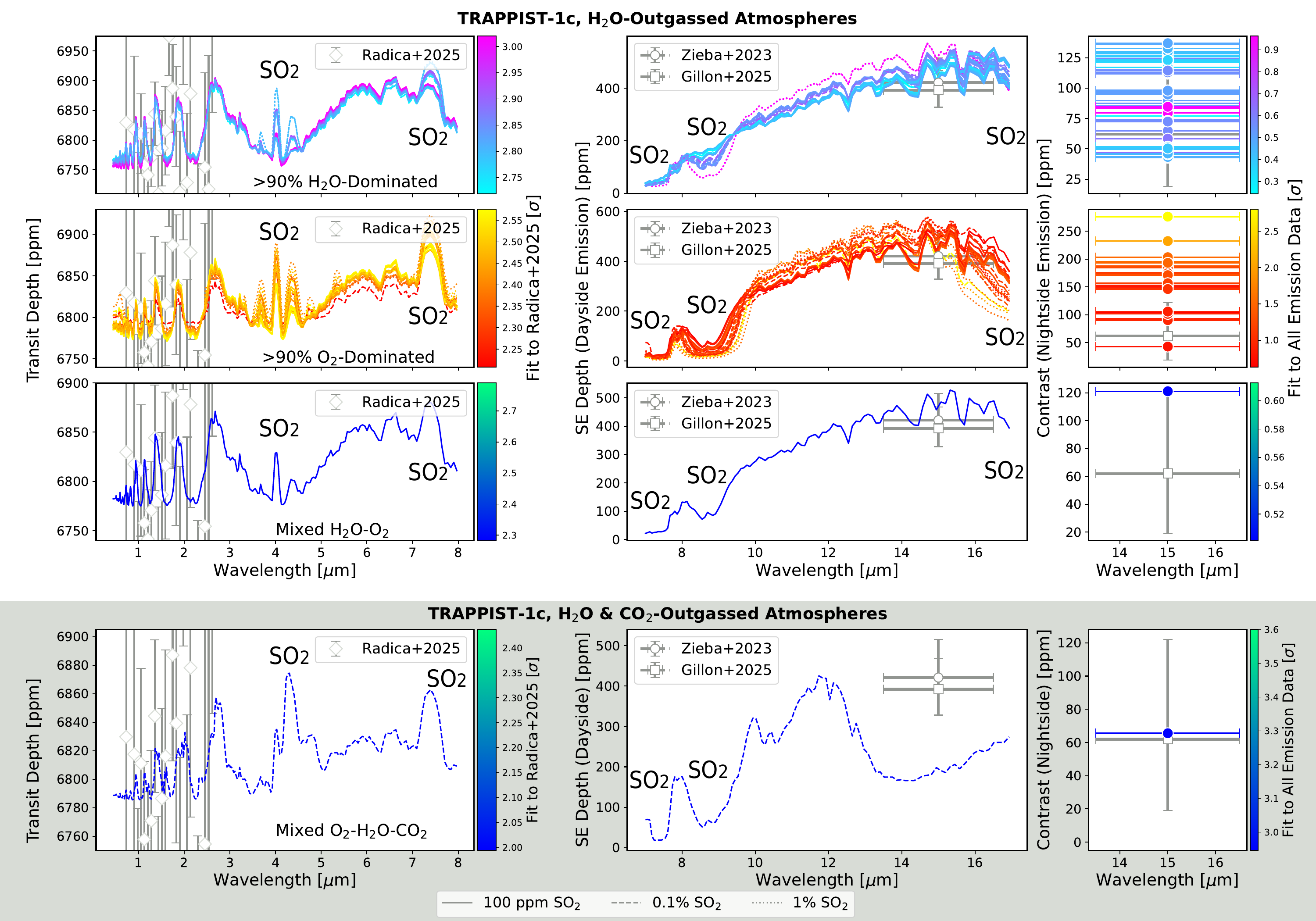}
    \caption{Same as Figure \ref{fig:T1cSpectraSummary}, but for the SO$_{2}$-bearing atmospheres of T-1c. Solid, dashed, and dotted lines indicate that 100 ppm, 0.1\%, or 1\% SO$_{2}$, respectively, was added to the bottom layer of the atmosphere. When including SO$_{2}$, no mixed CO$_{2}$-CO and H$_{2}$O-CO$_{2}$ outgassed O$_{2}$-dominated atmospheres were found to be stable, so those rows are not present in this figure. When including trace SO$_{2}$, all types of H$_{2}$O outgassed atmospheres may provide good ($<$1$\sigma$) fits to all available emission data.}
    \label{fig:SO2T1cSummarySpectra}
\end{figure}

\begin{figure}
    \centering
    \includegraphics[width=\linewidth]{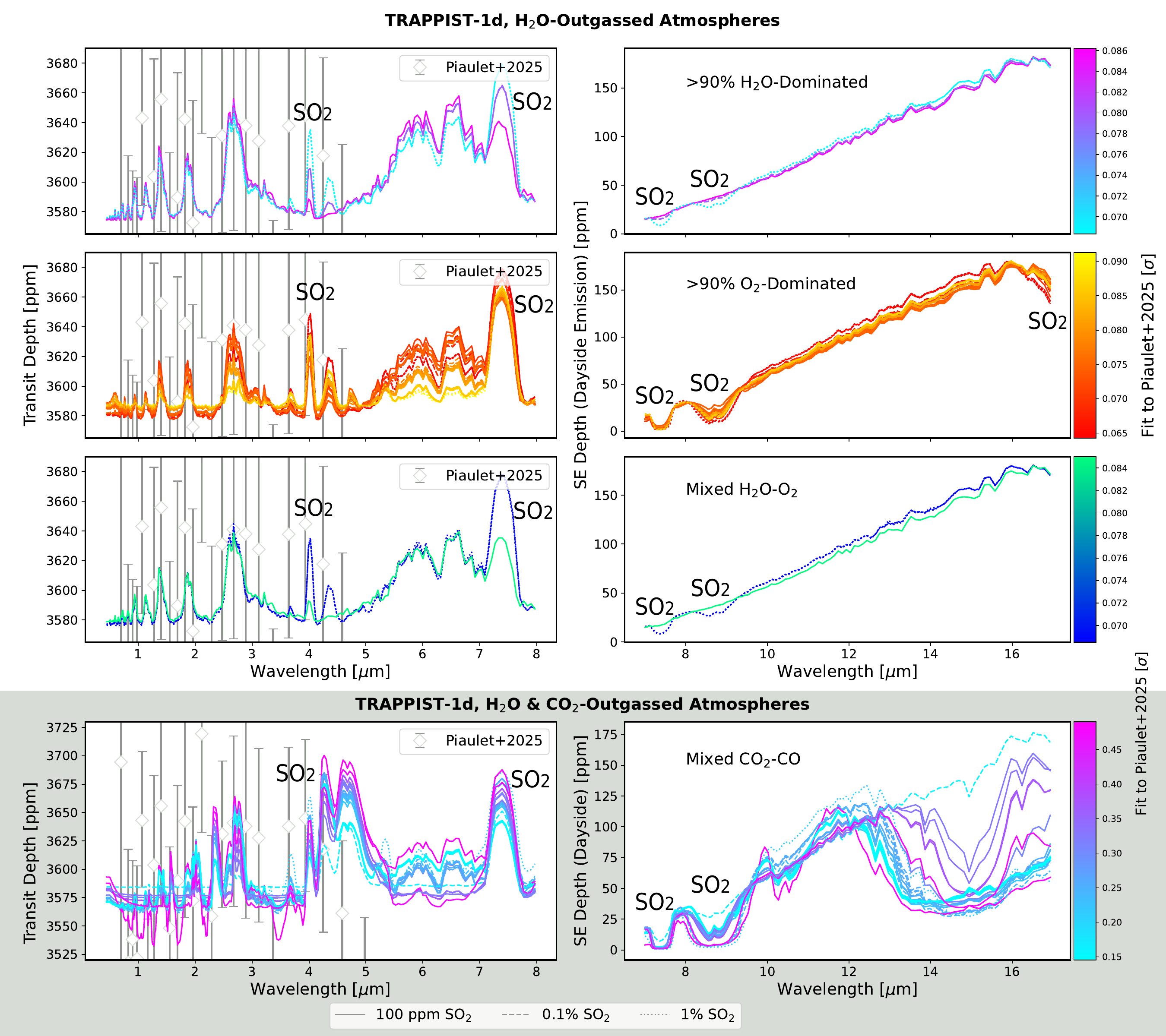}
    \caption{Same as Figure \ref{fig:T1cSpectraSummary}, but for the SO$_{2}$-bearing atmospheres of T-1d. Solid, dashed, and dotted lines indicate that 100 ppm, 0.1\%, or 1\% SO$_{2}$, respectively, was added to the bottom layer of the atmosphere. When including SO$_{2}$, H$_{2}$O-CO$_{2}$ outgassed O$_{2}$-dominated or mixed O$_{2}$-H$_{2}$O-CO$_{2}$ atmospheres were found to be stable, so those rows are not present in this figure. When including trace SO$_{2}$, the fit to available transmission data \citep[][]{Piaulet2025t1d} is changed negligibly, with all atmospheres providing fits to $<$1$\sigma$.}
    \label{fig:SO2T1dSummarySpectra}
\end{figure}

\begin{figure}
    \centering
    \includegraphics[width=\linewidth]{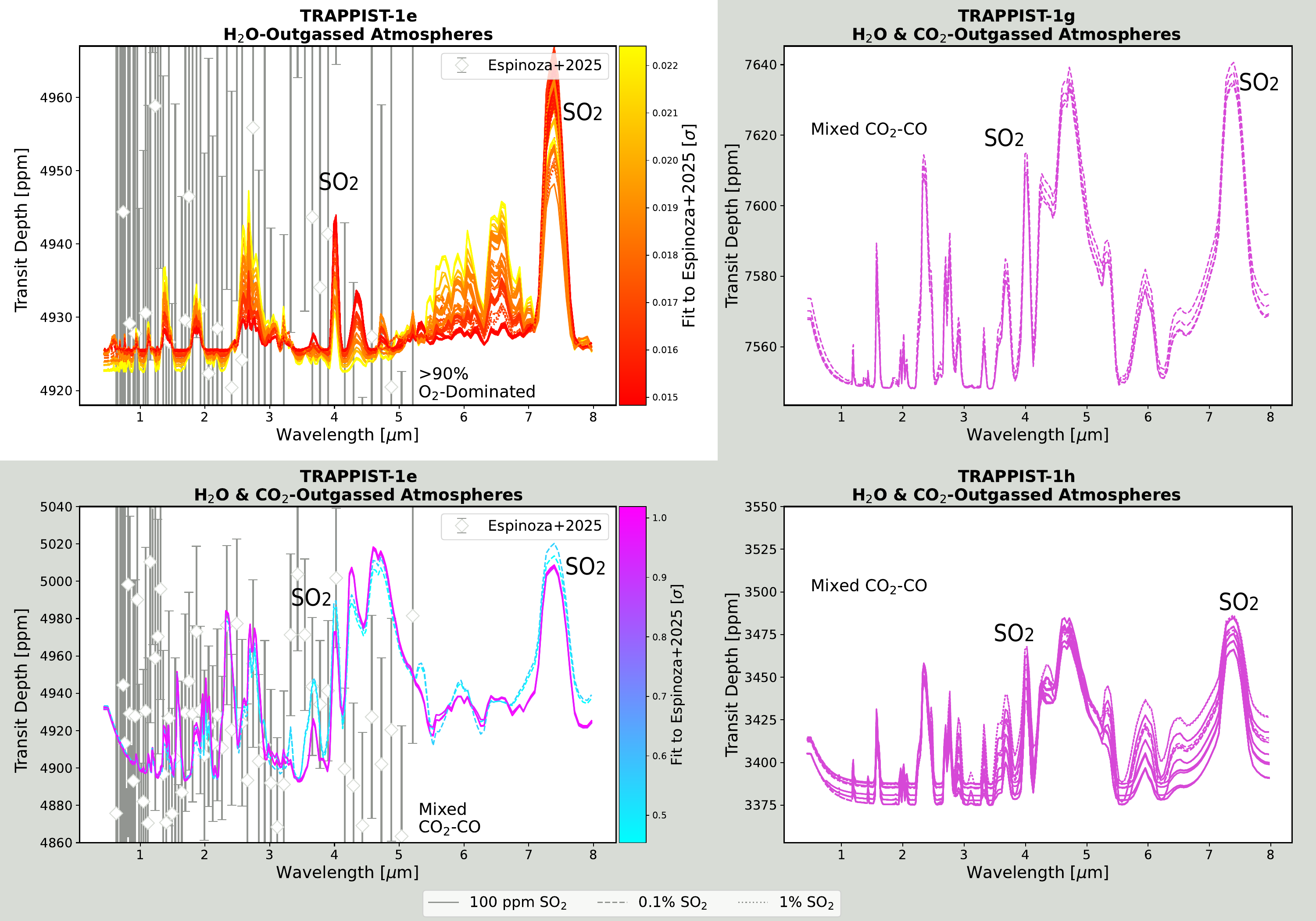}
    \caption{Similar to \ref{fig:T1cSpectraSummary}, but for the SO$_{2}$-bearing atmospheres of T-1e, g, and h. T-1f had no stable atmospheres containing SO$_{2}$ so it is not included here. The top and bottom left subplots show H$_{2}$O outgassed O$_{2}$-dominated and mixed CO$_{2}$-CO atmospheres for T-1e that contain trace SO$_{2}$; and the top and bottom right subplots show mixed CO$_{2}$-CO atmospheres with trace SO$_{2}$ for T-1g and h. Solid, dashed, and dotted lines indicate that 100 ppm, 0.1\%, or 1\% SO$_{2}$, respectively, was added to the bottom layer of the atmosphere.}
    \label{fig:SO2T1efghSummarySpectra}
\end{figure}

Beginning with T-1b (Figure \ref{fig:SO2T1bSpectraSummary}), the addition of trace SO$_{2}$ leads to stable, realistic atmospheres for only 2 compositions: mixed CO$_{2}$-CO and H$_{2}$O-CO$_{2}$ outgassed O$_{2}$-dominated. This is in contrast to the non-SO$_{2}$ results for T-1b, which also included O$_{2}$-dominated and mixed H$_{2}$O-O$_{2}$ atmospheres. While the H$_{2}$O only outgassed O$_{2}$-dominated and mixed H$_{2}$O-O$_{2}$ archetypes were stable without SO$_{2}$, no stable examples were found when trace SO$_{2}$ was included. As for the non-SO$_{2}$-bearing cases, these spectra all match the T-1b transmission data \citep[][]{lim2023atmospheric} to $<$1$\sigma$. However, all stable SO$_{2}$-bearing atmospheres for T-1b exhibited too much heat redistribution to match the available emission data.

When trace amounts of SO$_{2}$ were included for T-1c (Figure \ref{fig:SO2T1cSummarySpectra}), all atmospheric archetypes produced by H$_{2}$O alone were found to be stable but, when CO$_{2}$ outgassing was included, only stable mixed O$_{2}$-H$_{2}$O-CO$_{2}$ compositions were found, whereas the non-SO$_{2}$ results for T-1c also presented the other 2 possible CO$_{2}$-bearing compositions. 
The addition of SO$_{2}$ moderately improved the fit of several H$_{2}$O-dominated atmospheres in emission \citep[][]{Zieba2023t1cjwst,Gillon2025phase}, which all match to $<$1$\sigma$ in the SO$_{2}$-bearing cases. The fits of other atmospheric archetypes to the emission data, as well as fits to transmission data \citep[][]{Radica2025t1c} remain relatively unchanged with the addition of SO$_{2}$.  

For T-1d, when trace amounts of SO$_{2}$ were included (Figure \ref{fig:SO2T1dSummarySpectra}), all H$_{2}$O outgassed archetypes were stable, but only mixed CO$_{2}$-CO compositions were found when mixed H$_{2}$O-CO$_{2}$ outgassed atmospheres were considered. Compared to the non-SO$_{2}$ cases for T-1d, only the mixed O$_{2}$-H$_{2}$O-CO$_{2}$ archetype is missing when including SO$_{2}$. The fits to available T-1d transmission data \citep[][]{Piaulet2025t1d} for all atmospheres are improved when including SO$_{2}$, however, they remain well below 1$\sigma$ even before this SO$_{2}$ addition. 

The SO$_{2}$-bearing cases for T-1e, g, and h are shown on the same plot (Figure \ref{fig:SO2T1efghSummarySpectra}). Note, T-1f had no stable SO$_{2}$-bearing atmospheres in our modeling suite, so it is absent from these results. When trace amounts of SO$_{2}$ are included, mixed CO$_{2}$-CO atmospheres were stable for all of T-1e, g, and h. H$_{2}$O outgassed O$_{2}$-dominated atmospheres were also stable for T-1e. Fits to the T-1e transmission data \citep[][]{Espinoza2025t1e,Glidden2025t1e} remained relatively unchanged by the addition of SO$_{2}$ (all $<$1$\sigma$). Including trace amounts of SO$_{2}$ reduces the number of stable H$_{2}$O-CO$_{2}$ outgassed O$_{2}$-dominated and mixed O$_{2}$-H$_{2}$O-CO$_{2}$ atmospheres for both T-1e and g compared to cases without SO$_{2}$. 

\subsection{Data-Model Correlation} \label{subsec:CorrelationResults}

\begin{figure}
    \centering
    \includegraphics[width=0.8\linewidth]{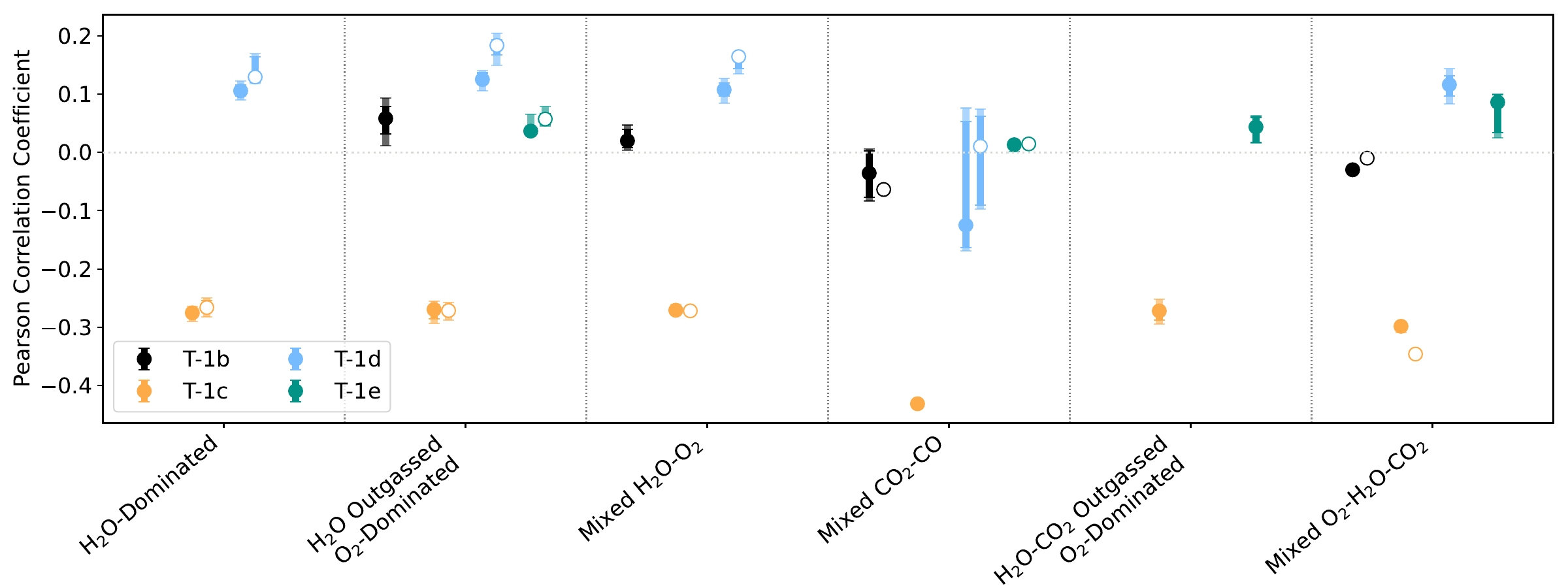}
    \caption{The 1 (darker colored bars) and 2$\sigma$ (lighter colored bars) ranges of Pearson Correlation Coefficients (PCCs) when comparing all atmospheres of a given archetype for a given planet to the available JWST transmission data for that planet (T-1b, c, d and e). The bars with filled and unfilled circular points denote atmospheres without and with the addition of trace SO$_{2}$, respectively. 
    The dotted horizontal grey line shows where PCC=0, points that fall above that line indicate a positive correlation and points that fall below indicate a negative correlation to the transmission data.}
    \label{fig:PCCsbcde}
\end{figure}

To further analyze and assess our data/model comparisons, we calculated Pearson Correlation Coefficients (PCC; Section \ref{subsubsec:CorrelationCoeffMethods}) for for all TRAPPIST-1 planets with observational data.  The PCC correlation provides a means to assess whether a given model predicts an increase or decrease in transmission depth at the same time that an increase or decrease is seen in a dataset. Specifically, absorption features in a spectrum are not independent, and a given molecule will provide a particular set of features in a spectrum. A positive PCC indicates that a model predicts absorption at the same wavelengths that potential absorption is seen in the data. 

The range of Pearson Correlation Coefficients (PCCs) are given by Figure \ref{fig:PCCsbcde}, which shows the 1 and 2$\sigma$ regions found from the distribution of PCCs across all stable atmospheres of a given archetype for a given planet. These results only pertain to the first 4 planets in the system (T-1b, c, d, and e) because the other planets do not have publicly available transmission data to compare to at the time of this paper. In Figure \ref{fig:PCCsbcde}, points with solid filled in circles correspond to the atmospheric archetype without the addition of trace SO$_{2}$ and points with unfilled circles include trace SO$_{2}$. 

Across our available models, correlation in either the positive or negative direction is generally weak ($<|$0.10$|$) for any atmosphere for planets T-1b, d, and e. Considering positive correlation as an indicator that the series of molecular features in the modeled transmission spectra may properly match the data, the atmospheric archetypes showing the strongest potential positive correlation to the T-1b data \citep[][]{lim2023atmospheric} are H$_{2}$O outgassed O$_{2}$-dominated atmospheres. For T-1c, the PCCs suggest a moderately strong ($\geq |$0.25$|$) negative correlation between the transmission data \citep[][]{Radica2025t1c} and all atmospheric archetypes found. Comparing to the T-1d data \citep[][]{Piaulet2025t1d}, all archetypes are generally positively correlated with equal strength ($\sim$0.1 -- 0.15), with the exception of mixed CO$_{2}$-CO atmospheres which skew slightly towards stronger negative correlation. Lastly, considering the T-1e data \citep[][]{Espinoza2025t1e,Glidden2025t1e}, all archetypes have negligible ($\sim$0.01) to weak ($\sim$0.1) positive correlation to the data, with the most positively correlated archetypes being H$_{2}$O outgassed O$_{2}$-dominated and mixed O$_{2}$-H$_{2}$O-CO$_{2}$. The addition of SO$_{2}$ does not appear to significantly affect the PCC values for T-1b, c, and e, marginally raising the PCC for T-1e H$_{2}$O outgassed O$_{2}$-dominated and T-1c mixed H$_{2}$O-O$_{2}$ atmospheres, but marginally lowering the PCC for T-1b mixed CO$_{2}$-CO atmospheres, as well as T-1b and T-1c mixed O$_{2}$-H$_{2}$O-CO$_{2}$ atmospheres. The addition of SO$_{2}$ does appear to raise the PCC values for all T-1d archetypes that produced stable SO$_{2}$-bearing models.

\section{Discussion} \label{sec:Discussion}

The expected high atmospheric escape rates driven by an M dwarf host star have raised concerns about the viability of thin atmospheres for planetary companions \citep[e.g.,][]{Dong2018trappistescape,Mansfield2024gl486b}. 
Using a self-consistent photochemical-climate model that also accounts for the balance of plausible atmospheric outgassing and escape rates, we have demonstrated that there is a broad range of stable tenuous atmospheres that could potentially exist on the TRAPPIST-1 planets. The stable atmospheres found in this study are \peerreview{constrained by} reasonable rates of either H$_{2}$O or mixed H$_{2}$O-CO$_{2}$ outgassing \citep[][]{Thomas2025outgassrates} and often exhibit thin (10 mbar -- 0.1 bar) surface pressures, between that of Earth and Mars. Despite the \peerreview{overall model tendency toward} low pressures, the compositions of these atmospheres vary widely, with a mix of compositions containing H$_{2}$O, O$_{2}$, CO$_{2}$, and CO. 
We have shown that plausible replenishment from outgassing may allow tenuous secondary atmospheres to be maintained, even when adopting escape rates for O, O$_{2}$ and CO$_{2}$ (up to $\sim$10$^{30}$ s$^{-1}$) that are many orders of magnitude higher than those seen for the Solar System terrestrials \citep[e.g., up to $\sim$10$^{25}$ s$^{-1}$ for neutral and/or ionized atomic O on Venus, Earth, and Mars;][]{Gronoff2020atmospheric}.

\peerreview{We begin by discussing the properties of atmospheres in our modeling suite (\S\ref{subsec:AtmospheresinStudyDiscuss}) including photochemistry and spectral features. We then compare our models to current observational datasets of the system (\S\ref{subsec:ModeltoDataDiscuss}) and consider how our atmospheric cases may be distinguished with future observations (\S\ref{subsec:fittingScenariosDiscussion}). We further consider the implications for traditional planetary habitability that arise from our pressure-temperature profiles (\S\ref{subsec:PTHabitabilityDiscuss}), the feasibility of escape fluxes adopted by our models following our treatment of effusion velocities (\S\ref{subsec:EscapeFluxFeasibleDiscussion}), and finally the limitations of our study (\S\ref{subsec:LimitationsDiscussion}).}

\subsection{Characteristics of Atmospheres in this Study} \label{subsec:AtmospheresinStudyDiscuss}

\peerreview{Here we discuss the physical properties of the atmospheres modeled in this study. First we begin with a discussion of the photochemical profiles (\S\ref{subsubsec:photochemdiscuss}), and then we describe the spectra seen across our atmospheres (\S\ref{subsubsec:spectradiscuss}).}

\subsubsection{\peerreview{Photochemical Profiles}}\label{subsubsec:photochemdiscuss}

The atmospheres in this study are maintained by 
outgassing, with vertical distributions and the generation of other gases governed by the destruction of outgassed H$_{2}$O, CO$_{2}$ and SO$_2$, and subsequent photolytic and inter-species reactions. 
For H$_2$O outgassing alone, the atmospheres found here are marked by efficient production of O$_{2}$, and to a lesser extent O and O$_{3}$, in the lower atmosphere. Post H$_2$O photolysis, the \peerreview{resulting} O$_{2}$ can build up to a significant fraction of the atmospheric composition due to the loss of H to space. In the upper atmosphere, a sharp increase in atomic O and ozone is facilitated by \peerreview{high} photolytic destruction of both H$_{2}$O and O$_{2}$. When including CO$_{2}$ outgassing, the general behavior of O, O$_{2}$, and O$_{3}$ remains similar, though production of these oxygen daughter species can also proceed via CO$_{2}$ \peerreview{photolysis}. In some cases water vapor is at sufficiently low abundance that CO$_{2}$ recombination is strongly inhibited \citep[][]{Gao2015stabilityCO2atms}, allowing CO (as well as O$_2$) to build up in the atmosphere, and even become the dominant constituent.  

We find that the addition of SO$_2$ tends to reduce the amount of atmospheric O$_{2}$ and H$_{2}$O, particularly in the lower atmosphere, and increase the amount of CO. On examination of our photochemical reactions and products, we find that the availability of SO$_x$ gases \peerreview{(e.g., SO, SO$_{2}$, SO$_{3}$)} provides a set of \peerreview{catalytic} reaction pathways that depletes O$_{2}$, often producing SO and O directly. 
Since SO$_{2}$ is more efficiently photolyzed in the upper atmosphere, yielding S and SO, both cases are seen where the addition of SO$_{2}$ can increase or decrease the upper atmospheric abundance of O$_{2}$. In the latter case, subsequent reactions of S with O$_2$ produce additional SO and atomic O, and reactions of O$_{2}$ with carbon contribute to increased CO (this case can be seen by T-1c's O$_{2}$-H$_{2}$O-CO$_{2}$ profile in Figure \ref{fig:CandEPhotochem}). 
Water may be depleted with SO$_{2}$ more indirectly than O$_{2}$, with various reactions containing sulfur-bearing species using up OH, as well as a direct reaction with SO$_{3}$ to produce H$_{2}$SO$_{4}$ (e.g., Figure \ref{fig:so2PTZPlot}). CO tends to be increased in atmospheres containing SO$_{2}$ due to several production reactions involving S, O$_{2}$, SO, HS, and HCO. 

We also found that the inclusion of SO$_{2}$ tends to increase ozone in atmospheres that do not contain CO$_{2}$ outgassing, and decrease it in those that do, \peerreview{compared to their baseline cases. For a water outgassed atmosphere, devoid of CO$_{2}$ and CO, the addition of SO$_{2}$ leads to increased destruction of OH via reactions with SO$_x$ gases, thus ozone increases --- as its primary destruction pathway with OH is suppressed. However, when CO$_{2}$ is also outgassed, OH is already suppressed by reactions with CO and consequently when SO$_{2}$ is added, it does not significantly change the availability of OH. Instead, SO$_x$ gases can deplete O$_{2}$, reducing the source of ozone, and it can even destroy ozone directly. This leads to an overall reduction of ozone in a joint H$_{2}$O and CO$_{2}$ outgassed atmosphere when SO$_{2}$ is injected.}


Considering inter-planetary comparisons, there are several notable results which appear to be a function of distance from the host star---including the stability of H$_{2}$O-generated atmospheres, and the strength of top-of-atmosphere escape. When considering all stable atmospheres per planet and comparing similar archetypes, the fraction of water vapor at the surface level of the atmosphere strictly decreases as distance from the host star increases.  This is due to the colder atmospheric temperatures overall on the outer planets, and increased condensation and precipitation of water out of the vapor stage as the temperature drops. The escape rate of O$_{2}$ also tends to be higher for interior planets due to the increased incident radiation available to drive escape (e.g., Figure \ref{fig:EscapeRates}). There is a similar, but weaker, distance dependence for the escape rates of atomic O and CO$_{2}$; it likely appears weaker because of the smaller mass of atomic O (easier to remove), and due to undersampling in the case of CO$_{2}$ (i.e., there were a relatively small number of stable cases that contained CO$_{2}$ to consider). 

In contrast to the globally-averaged pressure-temperature profiles for Solar System terrestrials, that exhibit cooling with altitude from the surface and an upper atmosphere with an isothermal profile or temperature inversion, several of our low surface pressure atmospheres displayed some unusual behavior.  For TRAPPIST-1 e in particular, several model atmospheres show heating with altitude in the lower atmosphere, followed by cooling with altitude in the upper atmosphere. This behavior is most prominently seen for O$_{2}$-dominated atmospheres generated by the H$_{2}$O-CO$_{2}$ outgassing,  and for mixed O$_{2}$-H$_{2}$O-CO$_{2}$ atmospheres (Figure \ref{fig:CandEPhotochem}). This behavior is driven by high atmospheric H$_{2}$O abundances in low surface pressure atmospheres ($\lesssim$10$^{-3}$ bars), which allows more rapid heating in the lower atmosphere facilitated by absorption from the H$_{2}$O (see bottom row Figure \ref{fig:CandEPhotochem}).

\subsubsection{\peerreview{Atmospheric Spectra}} \label{subsubsec:spectradiscuss}

Turning to spectral features, the three H$_{2}$O outgassed atmospheric archetypes in this work (prior to the addition of SO$_{2}$) all show spectra dominated by H$_2$O features\peerreview{, as this is the most active absorber in these atmospheres}. Atmospheres more strongly dominated by O$_{2}$ have lower water vapor abundance and display smaller water vapor features in transmission. \peerreview{This is similar in emission, where the spectral continuum is dampened by water vapor absorption from 12 -- 17 $\mu$m. All of the H$_{2}$O outgassed atmospheres provide adequate fits to transmission data for all planets ($\lesssim$3$\sigma$), as well as T-1c emission data, and the O$_{2}$-dominated atmospheres may provide $<$3$\sigma$ fits to the T-1b emission data.} For a given atmospheric pressure, adding H$_{2}$O tends to increase the nightside temperature and worsen the fit to the emission data \citep[][]{Gillon2025phase}. Similarly, even for O$_{2}$-dominated atmospheres, higher pressures also increases the nightside temperature and result in poorer fits \peerreview{(e.g., $>$2$\sigma$ for $>$0.4 bars on T-1c)}. Ozone features are relatively absent from the H$_{2}$O outgassed atmospheres, likely because reactions with OH (derived from H$_2$O photolysis) destroy O$_3$, and prevent it from building up to significant levels in the atmosphere.  However, weak O$_3$ features are seen in the O$_2$-dominated atmospheres where atmospheric water vapor is lower, and O$_3$ can persist and build up to higher abundances.  

The joint H$_{2}$O and CO$_{2}$ atmospheres have more diverse compositions and a larger collection of spectral features, including features from CO$_{2}$, H$_{2}$O, CO, and O$_{3}$. In all of the compositions containing CO$_{2}$-outgassing, the strong absorption features of CO$_{2}$ (e.g. near 2, 2.7, 4.3, 9.4, 10.4 and 15$\mu$m) can be seen with varying degrees of strength. Several smaller CO$_{2}$ features shortward of 1.5$\mu$m can also be seen, though they may be hard to disentangle from H$_{2}$O features in the same region. CO is primarily seen in the spectra of the mixed CO$_{2}$-CO composition atmosphere spectra at 2.35$\mu$m and 4.6$\mu$m. There is also a 4.7$\mu$m feature seen in the spectra of the H$_{2}$O-CO$_{2}$ outgassed O$_{2}$-dominated and mixed O$_{2}$-H$_{2}$O atmospheres.  However, in these compositions, it is caused by O$_{3}$ absorption as opposed to CO. The ozone Chappuis band additionally provides a relatively large feature spanning 0.5--0.7$\mu$m, though this is difficult to access with JWST, especially for planets orbiting M dwarfs, which produce relatively little radiation for transmission at these shorter wavelengths. 

When including SO$_{2}$ in the simulations, clear features appear primarily near 4$\mu$m, 7.3$\mu$m, and 8.7$\mu$m, and the general continuum in emission is increased between $\sim$10 -- 16$\mu$m and depressed at wavelengths longer than $\sim$16$\mu$m due to the SO$_{2}$ bending fundamental centered near 19.3$\mu$m.
In transmission, the clearest effect of including SO$_{2}$ is the presence of 3 absorption features centered about 4$\mu$m. 
These are particularly well delineated in the H$_{2}$O outgassed O$_{2}$-dominated atmospheres (e.g., T-1c Figure \ref{fig:SO2T1cSummarySpectra}, T-1e Figure \ref{fig:SO2T1efghSummarySpectra}). In emission, SO$_2$ features at 7.3$\mu$m and $\sim$8.7$\mu$m become saturated (caused by the stretching modes of SO$_{2}$), particularly in O$_{2}$-dominated atmospheres (see T-1c, Figure \ref{fig:SO2T1cSummarySpectra}). Furthermore, the overall emission continuum is raised (visible between $\sim$10$\mu$m to 16$\mu$m) compared to the non-SO$_{2}$ atmospheres. When CO$_{2}$ is present, the SO$_{2}$ absorption at $\sim$16$\mu$m is not discernible from CO$_{2}$. Addition of greenhouse gas SO$_{2}$ and subsequent photochemical reactions tends to increase both the day and nightside temperatures of planets, raising the continuum. This effect can contribute warming of up to 40 K on the dayside and 20 K on the nightside in the case of O$_{2}$-dominated atmospheres that have lower water abundances (Figure \ref{fig:PTDiagrams}). In steam atmospheres, whose compositions are already dominated by a strong greenhouse gas, the addition of SO$_{2}$ does not significantly change the temperature. 

\subsection{\peerreview{Model Comparisons to Current Data}} \label{subsec:ModeltoDataDiscuss}

When possible, we quantify the fit of atmospheres in this study to available JWST secondary eclipse, thermal phase curve and transmission data (using the $\chi^{2}$ goodness-of-fit test) and show that most current observations are not sufficient to distinguish, or rule out, the thin atmospheres we modeled to $>$3$\sigma$. Transmission data are currently available for T-1b \citep[][]{lim2023atmospheric}, c \citep[][]{Radica2025t1c}, d \citep[][]{Piaulet2025t1d}, and e \citep[][]{Espinoza2025t1e,Glidden2025t1e,Allen2025t1ePrelimResults}, and emission data are available for T-1b \citep[12.8 and 15$\mu$m;][]{Greene2023t1b,Ducrot2025combined,Gillon2025phase} and T-1c \citep[15$\mu$m;][]{Zieba2023t1cjwst,Gillon2025phase}. The transmission data for TRAPPIST-1 planets is now known to be thoroughly affected by stellar contamination \citep[][]{Rackham2018TLSE,lim2023atmospheric,Howard2023T1flaringJWST,Rathcke2025stellarcorrect}, which, in combination with small datasets, has led to large error bars on transmission points ($\pm$ $>$100 ppm) that exceed the size of even the largest spectral features for atmospheres in our study ($\lesssim$150 ppm). Due to this uncertainty versus feature strength size difference, transmission data is currently not sensitive to the tenuous atmospheres studied here. The lack of sensitivity at the $>$3$\sigma$ \secondpeerreview{level} to the tenuous ($<$0.1 bar) atmospheres in this work is in agreement with the conclusions reached by the studies presenting JWST transmission data \citep[][]{lim2023atmospheric,Radica2025t1c,Espinoza2025t1e,Glidden2025t1e}.

In emission, we find that only extremely thin ($\lesssim$0.01 bar) dry O$_{2}$ atmospheres match the available T-1b data \citep[][]{Greene2023t1b,Ducrot2025combined,Gillon2025phase} to $<$3$\sigma$, which is in agreement with past work suggesting a lack of significant greenhouse gases or thick atmosphere \citep[e.g.,][]{ih2023constrainingb}. Conversely, several compositions match T-1c emission data \citep[][]{Zieba2023t1cjwst,Gillon2025phase} to $<$1$\sigma$ including O$_{2}$-dominated atmospheres with no significant CO$_{2}$ at $\leq$0.3 bars or with $<$1\% CO$_{2}$ up to 0.05 bars, O$_{2}$-dominated with up to 15\% H$_{2}$O at $\leq$0.06 bars, or H$_{2}$O-dominated at $\leq$0.02 bars. Additionally, nearly every atmospheric scenario for T-1c presented here are consistent at $\leq$3$\sigma$ to the T-1c emission data \citep[][]{Zieba2023t1cjwst,Gillon2025phase}. This is in agreement with past work on analyzing these observations \citep[e.g.,][]{Lincowski2023t1catms} which found trace amounts of greenhouse gases (e.g., H$_{2}$O, CO$_{2}$) may be present in thin atmospheres or in atmospheres dominated by a less active constituent, such as O$_{2}$. 


After the comparisons with T-1e data given in \citet{Espinoza2025t1e,Glidden2025t1e}, additional preliminary results of a multi-cycle JWST program by \citet{Allen2025t1ePrelimResults} were published. However, these additional three T-1e transmission observations possessed comparable signal to data already available \citep[][]{Espinoza2025t1e,Glidden2025t1e} and did not change the confidence on distinguishing their best-fit atmospheric scenario from a flat line. Because the newly available data would not have changed the detection fits found in our results, we did not recalculate the $\chi^2$ goodness-of-fit metric for our models. Future data from the multi-cycle program for T-1e may change the detection fits of models in this work when signal is increased. Notably, \citet{Allen2025t1ePrelimResults} highlighted spectral features that may be attributed to CO, which will be tested with future observations in the multi-cycle program. \peerreview{While it is unclear if the potential CO absorption seen by \citet{Allen2025t1ePrelimResults} would be more likely caused by the planetary atmosphere or stellar photosphere,} our model simulations did generate several atmospheres rich in CO, which may be sustained both with and without trace SO$_{2}$ (Figures \ref{fig:T1efghSpectaSummary} and \ref{fig:SO2T1efghSummarySpectra}). These high CO atmospheres for T-1e are all between 0.5 -- 2$\sigma$ fits to the T-1e transmission data from \citet{Espinoza2025t1e,Glidden2025t1e}. 

\subsection{Distinguishing Atmospheric Scenarios \& Future Observing} \label{subsec:fittingScenariosDiscussion}

Even though many of the atmospheric spectra shown in Figures \ref{fig:T1cSpectraSummary} through \ref{fig:SO2T1efghSummarySpectra} are from 0.1 bar or even lower pressure atmospheres, they can still generate potentially observable 50-120 ppm features in transmission and emission spectra. The interior planets (T-1b, c, and d) in particular are hot enough that emission observations, which avoid the stellar contamination seen in transmission studies, can be use can be used to learn about their atmospheres. 
However, JWST data for T-1b and c are primarily in the 15$\mu$m bandpass \citep[][]{Greene2023t1b,Zieba2023t1cjwst,Ducrot2025combined,Gillon2025phase}, which is strongly sensitive to absorption from CO$_2$, but less compositionally diagnostic for atmospheres that don't contain it, which includes several of our O$_{2}$- and H$_2$O-dominated compositions. A drop in secondary eclipse depth, indicating a cooler dayside, can be due either to heat transport from the day to nightside by an atmosphere, or to CO$_2$ absorption, or a combination of both. To break this degeneracy for atmospheres that don't contain CO$_2$, absorption from additional diagnostic molecules, such as H$_2$O and O$_3$, needs to be sought. 

For O$_{2}$-dominated atmospheres produced by either H$_2$O or mixed H$_{2}$O-CO$_{2}$ outgassing, the and T-1b and T-1c results (Figure \ref{fig:T1cSpectraSummary}, \ref{fig:T1bSpectraSummary}, \ref{fig:SO2T1bSpectraSummary} and \ref{fig:SO2T1cSummarySpectra}) show that, beyond CO$_{2}$, the largest emission features are created when either trace O$_{3}$ or SO$_{2}$ is present at sufficient levels to create features at 10 and 8.7$\mu$m, respectively. In the case of ozone, MIRI's F1000W photometric filter may allow the 10$\mu$m bandpass to be accessed for a relatively small observing investment (verified with the JWST ETC\footnote{https://jwst.etc.stsci.edu/} and APT), but the width of this filter will prevent an observation from measuring the bottom of a deep 9.6$\mu$m ozone feature. To elaborate, the width of the F1000W filter is larger than that of the 9.6$\mu$m O$_{3}$ feature thus, an observation with this mode will include portions of the spectral continuum on either side of the feature and dilute the strength of the band and our sensitivity to it. For  SO$_{2}$, the broad 16 -- 22$\mu$m absorption feature may also be accessible with the F1800W, F2100W, or F2550W MIRI photometric filters, but the large 8.7$\mu$m absorption seen in emission is between two of MIRI's photometric filters (F770W and F1000W), so MIRI emission spectroscopy would be required. However, the large 7.3$\mu$m SO$_{2}$ feature may be accessible with the F770W photometric filter. 

Emission spectroscopy of the full 5 -- 14$\mu$m wavelength range, offered by MIRI LRS, would allow the observer to bin the data commensurate with the expected strength of the SO$_{2}$ 8.7$\mu$m feature, as well as the O$_{3}$ 9.6$\mu$m feature, simultaneously. Emission spectroscopy would also provide information on the strength of emission in the 10 -- 14$\mu$m range, which may additionally probe the presence of trace SO$_{2}$ amounts, which tends to increase emission at these wavelengths. An emission observing campaign of this kind would be especially valuable for T-1c, which has already demonstrated a lower eclipse depth at 15$\mu$m than expected for an airless body \citep[][]{Zieba2023t1cjwst}. T-1c additionally has 12.8$\mu$m unpublished MIRI photometric data. If this data showed increased emission at 12.8$\mu$m  compared to the 15$\mu$m measurement, this would suggest either the presence of SO$_{2}$ or CO$_{2}$, and follow-up observations targeting the 7.3 or 8.7$\mu$m SO$_{2}$ features would allow us to distinguish these two scenarios. 

There are several potentially detectable transmission features for H$_{2}$O, CO$_{2}$, CO, O$_{3}$, and SO$_{2}$ found in the atmospheres modeled in this work, but they are presently below the sensitivity of publicly available transmission data \citep[available for T-1b, c, d, and e;][]{lim2023atmospheric,Radica2025t1c,Piaulet2025t1d,Espinoza2025t1e,Glidden2025t1e}. Overall, the largest transmission features may be in the range of 75 -- 100 ppm, which is above the JWST noise floor of 5 -- 50 ppm \citep[depending on the instrument;][]{Rustamkulov2022nirspecnoisefloor,Lustig2023lhs475b,Bouwman2023MIRInoisefloor,Seager2025exoplanetsJWST}, and therefore, once stellar noise is better handled and sufficient observing time is invested, these larger features may be accessible. Beginning with H$_{2}$O, three clear features at 1.37, 1.88, and 2.7$\mu$m are within the wavelength ranges of both NIRISS SOSS \citep[used for T-1b and c;][]{lim2023atmospheric,Radica2025t1c} and NIRSpec PRISM \citep[used for T-1d and e;][]{Piaulet2025t1d,Espinoza2025t1e,Glidden2025t1e}, but these features are not detectable given the uncertainty on current data. Reducing this uncertainty could improve our ability to identify these H$_{2}$O features; however, H$_{2}$O is particularly problematic in transmission as it's also present in the stellar atmosphere \citep[e.g.,][]{Wakeford2018disentanglingh2otrappist}. H$_{2}$O also produces an observable feature at $\sim$6.3$\mu$m in many of our atmospheres, but this would only be possible to access with MIRI. Although preliminary studies with MIRI show it may need a factor of 10 or larger observing investment to detect water compared to NIRSpec \citep[][]{Lustig2019detectT1}, the now known challenges of stellar contamination may suggest that this is a worthwhile investment if water detection on the TRAPPIST-1 planets is a community priority. The large 4.3$\mu$m CO$_{2}$ feature (particularly seen in our mixed CO$_{2}$-CO and mixed O$_{2}$-H$_{2}$O-CO$_{2}$ archetypes) is accessible to both NIRISS and NIRSpec spectroscopy though, similar to H$_{2}$O, more observations are needed in addition to current data to be sensitive to the feature size in our tenuous atmospheres. Although CO and O$_{3}$ provide features at $\sim$4.6$\mu$m, these are also in the vicinity of strong CO$_{2}$ (4.3$\mu$m) and H$_{2}$O (5 -- 7$\mu$m) absorption, and so they may be more difficult to retrieve. SO$_{2}$ is potentially detectable in transmission due to a trio of absorption features centered at 4$\mu$m. Particularly, sharp absorption at 4.0 $\mu$m could show the presence of SO$_{2}$. 

Future transmission observations of the system will have to address the presence of stellar contamination by developing and refining strategies to reliably remove stellar contamination. \citet{Rathcke2025stellarcorrect} outlined a potential method for removing stellar contamination which involved observing back-to-back or overlapping transits of a target planet with T-1b, but this method relies on the assumption that T-1b is airless, which has been challenged by recent 12.8$\mu$m observations that indicated lower emission than expected \citep[][]{Ducrot2025combined}. Acquiring additional out-of-transit baseline in future observing campaigns may also help to provide context on stellar activity around transits, and provide new data to calibrate models for removing stellar contamination. Moreover, we demonstrated in \S\ref{subsec:CorrelationResults} how models may be positively or negatively correlated with transmission data. Given that the relative size of transmission features in our study are generally smaller than the size of uncertainty on data points, it would be expected that this correlation coefficient should be close to zero. While this is the case for some scenarios, T-1c in particular shows a strong negative correlation ($\sim$ -0.3) for all compositions, with dips below the spectral continuum coincident with the locations of multiple water bands. This may indicate an over-removal of expected water contamination from the stellar atmosphere.

\subsubsection{Implications for the Rocky Worlds DDT Time}\label{subsubsec:DDTTimeDiscuss}

The tenuous, outgassing sourced atmospheres and their spectral discriminants generated in this work may be broadly applicable to interpretation of observations other M dwarf terrestrial planets, if they are expected to follow similar evolutionary paths to those in the TRAPPIST-1 system.  The joint JWST and Hubble Space Telescope (HST) Rocky Worlds DDT program \citep[][]{DDTwebsite,Redfield2024DDTTime} will use MIRI photometry to observe the 15$\mu$m band of up to a dozen terrestrial exoplanets that orbit M dwarf stars. The archetype atmospheres in our work predict various absorption strength at the 15$\mu$m bandpass, ranging from strong in atmospheres containing significant CO$_{2}$, to weak, particularly in dry O$_{2}$-dominated atmospheres. Based on the 15$\mu$m data from the Rocky Worlds program, this work may be used to inform follow-up observations. Furthermore, we provide all of the temperature and mixing ratio profiles, as well as atmospheric spectra, generated through this work on GitHub\footnote{https://github.com/mgialluca/T1Atmospheres\_Gialluca2026}. Spectra could quickly be recreated to represent Rocky Worlds targets given an alternative host star spectrum and planetary parameters using the temperature and mixing ratio profiles provided here.

\subsection{Pressure-Temperature Profiles \& Habitability} \label{subsec:PTHabitabilityDiscuss}

Observations of T-1d could provide an excellent test of the habitable zone concept \citep{Kopparapu2013hz}, as it sits within TRAPPIST-1's optimistic habitable zone, but closer to the star than the conservative habitable zone's inner limit \citep[][]{Gillon2017seven,Kopparapu2013hz}.  Existing simulations of T-1d's possible climates have not converged on whether or not it is likely to be habitable, 
although many suggest either runaway states, or surface temperatures very much greater than 300 K, depending on atmospheric composition \citep[][]{Wolf2017climate3dtrappist,Lincowski2018trappist,Turbet2018T1ClimateDiversity,Turbet2020reviewpossibleT1Atms,Meadows2023t1biosigs}.  However, a small subset of the existing simulations---which were performed for 1 bar or greater pressure atmospheres---were found to, at least for the current insolation, produce marginally habitable surface temperatures. These cases included  dry O$_{2}$-dominated atmospheres \citep[][]{Lincowski2018trappist}, or more modern-Earth-like scenarios, where the presence of reflective water clouds contributed to cooling \citep[e.g.,][]{Turbet2018T1ClimateDiversity,Meadows2023t1biosigs}. 
In contrast to previous simulations, the majority of our T-1d models sustain potentially Earth-like temperatures (273.15 -- 300 K), but for atmospheric pressures that are $<1$ bar. Our lower pressure atmospheres produce habitable temperatures even for atmospheres containing greenhouse gases such as our \peerreview{H$_{2}$O-dominated atmospheres, as well as our clear-sky CO$_{2}$/CO, although we note that these latter atmospheres are more likely to be desiccated. Overall these} thinner atmospheres are conducive to cooling, and predict that surface temperatures for TRAPPIST-1 d will stay closer to its equilibrium temperature ($\sim$280 K for 0 albedo, $\sim$260 K for an albedo of 0.3). \peerreview{For the higher pressure atmospheres ($\sim$1 bar) the dayside temperatures are seldom habitable, but habitable conditions are seen on the night side, especially for the mixed O$_2$-H$_2$O-CO$_2$ atmospheres, which are relatively Earth-like in composition.} 


The atmospheres our model found to be stable for T-1e are also quite thin ($<$1 bar), which leads to either colder climates that mostly fall within the `habitable snowball' (242 -- 273.15 K) temperature range, but with a smaller number in the  Earth-like temperature regimes (273.15 -- 300 K). In contrast, many previous studies find the Earth-like temperature regime to be common for T-1e models \citep[e.g.,][]{Wolf2017climate3dtrappist,Lincowski2018trappist,Fauchez2020thai,Sergeev2022thaiT1e,Meadows2023t1biosigs}, although this can require the addition of greenhouse gas abundances, e.g. for CO$_2$ of several percent \citep[e.g.,][]{Meadows2023t1biosigs}. However, even in our models producing potentially `habitable snowball' temperatures, many of our T-1e atmospheres possess pressures too low to exit the `gas' phase of the water phase diagram (Figure \ref{fig:PTDiagrams}). The compositions that fall squarely within the liquid region of the water phase diagram for T-1e are those atmospheres rich in the greenhouse gas CO$_{2}$, and  with trace amounts of SO$_{2}$. \citet{Turbet2022thaiDryPlanets} also reported colder to `habitable snowball' global mean surface temperatures for dry planetary cases applied to T-1e ($\sim$215 -- 245 K), which is similar to many of our models. 

It is worth noting that the recent study by \citet{Allen2025t1ePrelimResults} on JWST transmission data for T-1e highlighted the potential presence of CO in the atmospheric spectra. In this study, the mixed CO$_{2}$-CO atmospheres (the only case we show with detectable CO) provide the warmest surface temperatures for T-1e in our suite, particularly when also including trace SO$_{2}$. However, we find that nearly all of the T-1e models in our study fit the early transmission data for TRAPPIST-1 e \citep[][]{Espinoza2025t1e,Glidden2025t1e} to $\leq$1$\sigma$ (Figures \ref{fig:T1efghSpectaSummary} and \ref{fig:SO2T1efghSummarySpectra}). Consequently the data are currently unable to distinguish between our model predictions, but model outcomes may be distinguished or ruled out as further transmission data becomes available.
Finally, we note that there are several cases where the globally-averaged nightside temperatures of models for T-1b and c fall within the liquid regime of the water phase diagram (Figure \ref{fig:PTDiagrams}) and the Earth-like temperature region (273.15 -- 300 K). Although the model used in this work has insufficient spatial resolution to map temperatures across the nightside \peerreview{and explicitly determine the nightside temperature minimum, which governs whether or not bulk components of the atmosphere can condense and collapse the atmosphere, the compositions modeled here are unlikely to collapse in these pressure-temperature regimes. For example, T-1c atmospheres in the liquid water regime have globally-averaged nightside temperatures between 250 and 550K, which is well above the $<$90K required to condense and collapse an O$_2$-dominated atmosphere \citep{Fray2009,Turbet2018T1ClimateDiversity}. 
We also note that many of the atmospheres plotted in Figure \ref{fig:PTDiagrams} may have localized regions where liquid water is possible, as their day and night conditions span the liquid water region of the diagram.}     

\peerreview{For T-1f, g, and h} we have modeled precipitation of water as ice, and find that when water is outgassed alone it is frozen before it can be photolyzed to produce O$_{2}$; \peerreview{however, if CO$_{2}$, an additional greenhouse gas, is outgassed along with water vapor, this can limit water ice formation. It is also worth noting that we began our simulations only with thin (0.1 bar) pressures of O$_{2}$, excluding any significant greenhouse gas --- this was with the intention of testing whether secondary atmospheres can be fully generated from outgassing.  While \citet{Turbet2018T1ClimateDiversity,Turbet2020reviewpossibleT1Atms} showed that CO$_{2}$ may freeze out on the outer TRAPPIST-1 planets, sufficient partial pressures of N$_{2}$ could help to prevent this.} 

\subsection{Escape Flux Feasibility} \label{subsec:EscapeFluxFeasibleDiscussion}

An important caveat when interpreting results dependent on escape rates is that our atmospheric model does not consider ionization or ionized species, although this may not strongly impact our results, as neutral escape processes may still dominate.  Non-thermal escape processes relying on the loss of ions (such as O$^{+}$, O$_{2}^{+}$, and CO$_{2}^{+}$) are expected to be quite high for rocky M-dwarf planets, such as those in the TRAPPIST-1 system \citep[e.g.][]{Li2025ionescapeKepler,Dong2018trappistescape}. However, neutral escape processes are also expected to be considerable (e.g., Jeans, photochemical, etc) \citep[][]{Gronoff2020atmospheric}. For H$_{2}$ and H specifically, a diffusion-limited escape flux is calculated more rigorously, following previous work with the \textit{Atmos} photochemical model \citep[e.g.][]{Arney2016pods,Lincowski2018trappist}. For our other species, we do not consider ionization, or model the detailed physics of TOA escape, we instead approximate total escape processes with one TOA flux out of the atmosphere for any given escaping species considered (O, O$_{2}$, CO$_{2}$). Total TOA escape fluxes are found by combining a fixed effusion velocity with the number density of the escaping species in the top layer of the atmosphere (detailed in the Methods section, \S\ref{sec:methods}). While the escape rates found in the present study are higher than those in \citet{Dong2018trappistescape}, their MHD model captures only ion escape channels, whereas our effusion fluxes more closely represent neutral escape channels. Brain \& Hinton (\textit{priv. comm.}) find neutral escape from an exo-Mars (CO$_2$-dominated) around Barnard's Star (M dwarf) to be 1 -- 3 orders of magnitude greater than ion escape. The findings of the present study show a similar trend for O$_{2}$ dominated atmospheres where our TOA escape flux (assumed to include neutral escape) may be up to 3 orders of magnitude greater than the pure ion escape rates for TRAPPIST-1 planets found by \citet{Dong2018trappistescape}.   

The majority of our T-1b and c atmospheres that best fit emission data display oxygen escape rates that are relatively high ($\sim$10$^{27}$ -- 10$^{30}$ s$^{-1}$) when compared to other values in the literature for TRAPPIST-1 ion escape \citep[][]{Dong2018trappistescape} or solar system terrestrials \citep[][]{Gronoff2020atmospheric}, but this may be realistic. As illustrated by the grey regions in Figure \ref{fig:EscapeRates}, all of the T-1b and c models identified as $<$3$\sigma$ fits to the available emission data possess O$_{2}$ escape rates 1 -- 3 orders of magnitude larger than the upper limit on O$_{2}^{+}$ escape rates provided by \citet{Dong2018trappistescape} magnetohydrogynamic simulations. However, \citet{Dong2018trappistescape} simulated  O$_{2}^{+}$ escape from Venus-like (CO$_{2}$-dominated) atmospheres, and in our O$_2$-dominated atmospheres there is a higher concentration of O$_2$ available for escape at the exobase, increasing escape rates. The \citet{Dong2018trappistescape} results also focus on ion escape, with reported expected escape rates on the order of 10$^{26}$ -- 10$^{27}$ s$^{-1}$ for O$^{+}$, and 10$^{25}$ -- 10$^{26}$ s$^{-1}$ for both O$_{2}^{+}$, and CO$_{2}^{+}$.  In contrast, our higher rates account for escape of neutral species, and as discussed above, neutral escape for planets orbiting M dwarfs, even for CO$_2$ atmospheres that likely have reduced oxygen escape rates, can be 1 -- 3 orders of magnitude greater than that due to ion escape (Brain \& Hinton, \textit{priv. comm.}), suggesting that the literature's simulated ion escape and our neutral escape rates are not inconsistent with each other. 


In contrast to solar system bodies, which all generally have higher O escape than O$_{2}$ at present-day \citep[][]{Gronoff2020atmospheric}, the majority of our simulated atmospheres ($\sim$68\%) exhibit higher O$_{2}$ escape rates than those for O. A possible explanation may be that the vast majority of stable atmospheres in this work are O$_{2}$-dominated ($\sim$58\%), so the larger concentration of O$_{2}$, when compared to O, may allow for higher escape rates. Furthermore there is some precedent in the solar system for O$_{2}$ loss rates to exceed that of O, occurring during certain events and time periods in the Martian atmosphere when the planet experienced heightened EUV due to solar winds or activity \citep[][]{Ma2004marslossmhd,Ma2018marsloss}. 

\subsection{Study Limitations \secondpeerreview{\& Directions for Future Work}} \label{subsec:LimitationsDiscussion}


In this work, we have provided an initial exploration of where the balance of photochemical, climate, outgassing, and atmospheric loss processes may support tenuous secondary atmospheres on the TRAPPIST-1 planets; however, there are several limitations of this work that should be noted when interpreting the results shown here, most notably due to our incomplete sampling of possible outcomes, and the limited number of outgassed species considered.  Importantly, we have not explored a statistically sized sample of the phase space defined by the ranges of outgassing and escape processes. Because of this sampling limitation, we cannot quantify statistical likelihoods that the planets possess an atmosphere, or make any predictions as to their most likely compositions. 
We can instead say that we have identified cases where specific atmospheres and compositional archetypes are plausible within the phase space. However, there may be boundary combinations that we did not test that could lead to compositions that are missing in our results for particular planets. Input combinations that were not explored also have the potential to lead to spectral features of varying strength and detectability compared to the sample of spectra given here. For example, an O$_{2}$-dominated atmosphere that has less water than the sample set in this work could lead to a larger ozone contribution and a deeper, and potentially more observable, feature at 9.6$\mu$m.
\secondpeerreview{It should also be noted that, while we specifically tested the generation and subsequent maintenance of a secondary atmosphere from a nearly airless state (0.1 bars O$_{2}$), it is also possible for M dwarf planets to retain thicker remnants of a primordial atmosphere at present-day, depending on initial volatile content \citep[e.g.,][]{Kite2020exoplanetatmlossrevive,Gialluca2024implications}. Given the tenuous atmosphere or airless conditions likely for the TRAPPIST-1 planets implied by observation \citep[e.g.,][]{Greene2023t1b,Zieba2023t1cjwst,Gillon2025phase}, and the high expected atmospheric escape \citep[e.g.,][]{Dong2018trappistescape}, we focused this study on atmospheres that could be sustained by the balance of geochemical outgassing and escape, without the need for preserved primordial material. However, a similar study could be conducted in the future with initial atmospheric states derived from the remnant of a primordial atmosphere. }

One particular difficulty in applying a statistical sampling method to a study of this nature is the computation time required by each of the model components, particularly climate. As a part of one \textit{VPL Climate} simulation, a full line-by-line radiative transfer solution is calculated around 2 -- 3$\times$ on average, and a climate simulation may need to be run many times before it is converged. In total, some atmospheres may take multiple days to achieve global convergence. Future work could explore ways to improve computational efficiency, and provide a more comprehensive sample of the diversity of plausible atmospheres.  

In the particular case of T-1f, which had a notably lower number of stable atmospheres than the other planets in this study (Table \ref{tab:NumberOfAtms}), many of the atmospheric input combinations tested failed to achieve global convergence. This is a state when, on a single pass through all of the photochemical, surface pressure, and climate models, the climate model finds local convergence on the first attempt and the surface pressure never changes by more than 3\%. In the case of T-1f, many simulations were not able to find local climate convergence on the first attempt in one pass through the global model. Our convergence strategy was very strict, as oscillations in the temperature profile of the uppermost atmospheric layers can prevent local climate convergence, even when the bulk atmosphere is not changing considerably. Future work could explore a more sophisticated scheme for convergence checking within the climate model that takes each individual atmospheric layer into consideration, as opposed to the global average used in this project. A more advanced convergence method may identify a larger sample of stable atmospheres for all planets in this study. 

Another limitation to note is that the array of gases we considered as outgassed or escaping species (H$_{2}$O, CO$_{2}$, and their daughter products) is likely a subset of the species that could be present. For example, from solar system terrestrials we know other constituents may be outgassed including CH$_{4}$, sulfur-bearing species, or chlorine-bearing species \citep[e.g.,][]{Gaillard2021outgassing,Redwing2022ClonIo}. There may be many more possible compositions for these planetary atmospheres  when considering more complex outgassing compositions, that were not explored in the present study. Similarly, our treatment of atmospheric sinks (e.g., deposition and escape) was simple, assuming some flux out of the bottom or top of the atmosphere. Future work may look into coupling our simulations to more advanced models of escape or surface-atmosphere exchange. 
Finally, while we did consider a separated day and night side climate calculation for the interior planets, we did not have the capability to model water condensation and atmospheric collapse on the nightside that may be expected on tidally locked planets and could be another pathway to complete atmospheric erosion \citep[][]{Turbet2018T1ClimateDiversity,Turbet2020reviewpossibleT1Atms}.

\peerreview{Clouds, dust, and hazes also constitute an area of potential uncertainty, as these influence  planetary climate and can modify atmospheric spectra. However, due to the low surface pressures and dry compositions seen across the majority of our models, clouds may only have a minimal effect on the spectra presented here. In general, at the low pressures exhibited by the majority of our stable atmosphere outcomes, diffusive and convective mixing become weaker, which are necessary processes for gas replenishment and cloud growth \citep[][]{Helling2019exoplanetcloudreview}, suggesting that significant cloud formation is unlikely in these atmospheres. In terms of Earth-like water clouds, we tracked the condensation of H$_{2}$O in our models and found no evidence for this on T-1b through d.  H$_{2}$SO$_{4}$ also does not condense on T-1b \citep[][]{Lincowski2018trappist}. Water did condense on T-1e and the outer planets in the system, but for T-1f through h this often led directly to permanent loss of water from the atmosphere, and atmospheric collapse in the H$_{2}$O only outgassing cases. Thus, T-1e may have the highest potential in our model to form water clouds.  However, even if clouds do successfully form, they may have very little impact on the predicted transmission spectra for many of our modeled outcomes. Looking at the formation of sulfuric acid clouds in Venus-Like high pressure CO$_{2}$-atmospheres for the TRAPPIST-1 planets, \citet{Lustig2019mirage} showed that H$_{2}$SO$_{4}$ clouds would truncate T-1c and d transmission at altitudes near 0.01-0.001 bars and T-1e transmission spectra at about 0.01 bars, which has the potential to affect the smaller fraction of our model outcomes that have higher surface pressures than this. However, given that  sulfuric acid cloud formation occurs at lower altitudes for planets further from the star, \citet{Lustig2019mirage} showed that the clouds formed at pressures above the limiting opacity of the atmosphere itself for planets beyond the orbit of T-1e, such that the clouds were essentially being invisible, which produced a strong match between their cloudy and clear-sky spectra \citep[][]{Lustig2019mirage}. They also showed that emission spectra may be more robust against influence from clouds than transmission observations \citep[][]{Lustig2019mirage}. In general, while the addition of clouds may perturb the climate and/or transmission spectra of a subset of the higher surface pressure atmospheres in this study, we have focused on presenting valid clear-sky solutions for the lower pressure atmospheres that are so far a better fit to T-1b and c data \citep[][]{Greene2023t1b,Zieba2023t1cjwst,ih2023constrainingb,Lincowski2023t1catms}, and we will leave cloud modeling for future work.}

\peerreview{Dust and hazes are likely more influential than clouds for the low pressure regime of many of our atmospheric models. Dust in particular provides a large source of uncertainty, as comparisons with Mars indicate that absorption features caused by dust can overlap spectroscopically with the regions of interest in this work \citep[][]{Erard1994martianDust,Bandfield2003carbonateMarsDust}. Dust may be more important for reflected light spectra, which we do not present here, but will be of interest for the upcoming Habitable Worlds Observatory. Hazes additionally may be present on these planets, but the formation of organic hazes requires the presence of methane \citep[][]{Arney2016pods,Turbet2020reviewpossibleT1Atms} which is not considered in our atmospheric models. Hazy methane atmospheres may also be valid solutions for the TRAPPIST-1 planets, but are not the focus of this study. Future work could consider the effect of dust, hazes, and clouds on climate and atmospheric spectra for tenuous atmospheres on the TRAPPIST-1 planets.}

\section{Conclusions} \label{sec:Conclusions}

We have conducted self-consistent photochemical-climate calculations for terrestrial exoplanets orbiting TRAPPIST-1, across a range of outgassing and escape processes. We show that the high expected atmospheric escape rates do not preclude tenuous atmospheres that may be consistent with current JWST data. Our model is the first to couple the processes of climate, photochemistry, outgassing and escape, and account for variable surface pressures as these atmospheres evolve towards model equilibrium. Through plausible water and/or CO$_{2}$ outgassing rates constrained for these planets by \citet{Thomas2025outgassrates}, modified by photochemistry and balanced by escape, we demonstrate that an array of atmospheric compositions may arise in the system including steam, dry or wet O$_{2}$, mixed CO$_{2}$-CO, and mixed O$_{2}$-H$_{2}$O-CO$_{2}$. Most commonly, our stable atmospheres have 0.1 mbar -- 1 bar surface pressures, and exhibit a range of surface temperatures from 100 -- 580K. For T-1 d and e, we find several stable atmospheres with globally-averaged temperature and pressure combinations that permit liquid surface water. We also find similar potentially habitable conditions for the hemispherically-averaged nightsides of TRAPPIST-1 b and c. 


Our atmospheres present relatively large spectral features in transmission, even for some 0.05 bar terrestrial atmospheres ($\lesssim$150 ppm) but are still typically smaller than the uncertainty associated with JWST transmission data from the system to date \citep[$\pm >$100 ppm;][]{lim2023atmospheric,Radica2025t1c,Espinoza2025t1e,Glidden2025t1e,Allen2025t1ePrelimResults}, indicating that we do not currently have the sensitivity to distinguish or rule out these atmospheres to reasonable confidence ($>$3$\sigma$) in transmission. Conversely, emission data available for T-1b and c provides more robust measurements, though limited to only one or two wavelengths (12.8 or 15$\mu$m). Comparison to emission observations suggests that extremely thin ($\sim$0.01 bar) dry O$_{2}$-dominated compositions without significant CO$_{2}$ may provide the best atmospheric match to the current emission data for T-1b \citep[$<$3$\sigma$;][]{Greene2023t1b,Ducrot2025combined,Gillon2025phase}. In contrast, the best atmospheric fits found for the T-1c emission data \citep[$<$1$\sigma$;][]{Zieba2023t1cjwst,Gillon2025phase} are from slightly thicker O$_{2}$-dominated atmospheres ($\leq$0.3 bars) with trace fractions of greenhouse gases, though higher fractions of greenhouse gases are equally plausible if the atmospheric pressures are lower. 
Given the low pressures in our models, even some atmospheres with up to $\sim$20\% CO$_{2}$ cannot be ruled out unambiguously by 15$\mu$m data alone for T-1c if sufficient water ($\sim$10\%) is also present to heat the atmosphere.

%

%



\section{Acknowledgements}

We would like to thank Josh Krissansen-Totton for helpful discussion on dry deposition. We also thank two anonymous peer reviewers for their comments which greatly improved the quality of our manuscript. 
MTG acknowledges funding from the NSF Graduate Research Fellowship Program (DGE-2140004). MTG, VSM, APL, and TBT were supported in part by the NASA Virtual Planetary Laboratory Team, a member of the NASA Nexus for Exoplanet System Science, and funded via NASA Astrobiology Program Grants 80NSSC18K0829 and 80NSSC23K1398. PCH and DAB were supported by the Retention of Habitable Atmospheres in Planetary Systems grant 80NSSC23K1358 awarded by the NASA ICAR program. This work made use of the advanced computational, storage, and networking infrastructure provided by the Hyak supercomputer system, which was funded by the University of Washington and the NASA Virtual Planetary Laboratory.

\appendix

\renewcommand\thefigure{\thesection.\arabic{figure}}
\setcounter{figure}{0}





\section{Correlation Coefficient Proof of Concept} \label{AppendixSec:PCCProof}

\begin{figure}
    \centering
    \includegraphics[width=0.5\linewidth]{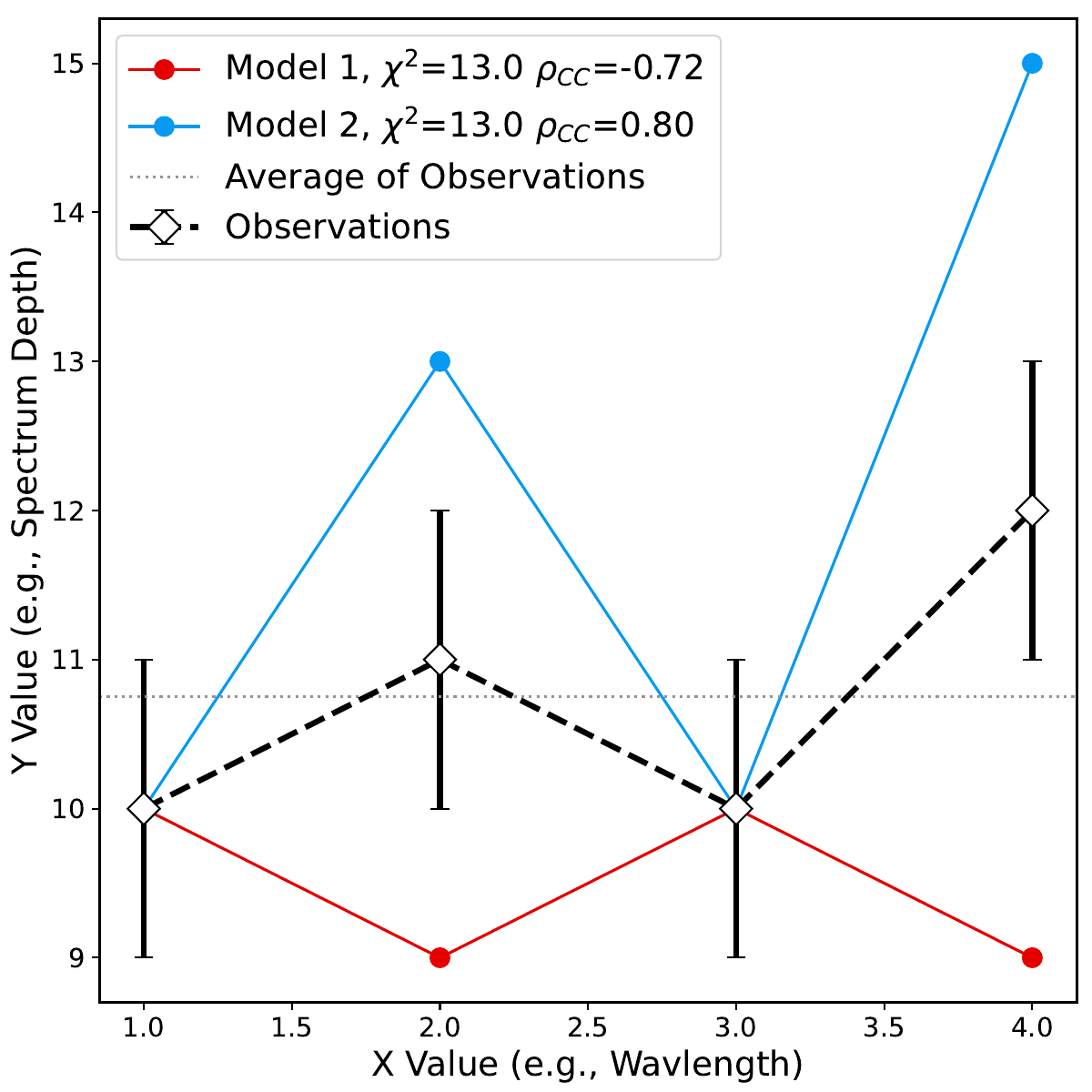}
    \caption{An example dataset (black points connected by dashed line) compared to two models (red and blue lines). The average of the observations is shown by the dotted grey line to allow better by-eye comparisons. While the models both display the same $\chi^{2}$ fit (13 in this case), model 1 and 2 are strongly negatively and positively correlated, respectively, suggesting model 2 may be a better overall fit if this example is taken to represent transmission spectra.}
    \label{fig:correlationcoeff}
\end{figure}

Here we provide a simple example that illustrates how a model can be preferred over another using the Pearson Correlation Coefficient (PCC) when they both report the same $\chi^{2}$ value. Figure \ref{fig:correlationcoeff} illustrates the example we will describe here. Consider a set of observations with associated uncertainty (black diamond points connected by dashed lines, Figure \ref{fig:correlationcoeff}), which could be transmission points at specific wavelengths. We want to compare two different models (blue and red lines, figure \ref{fig:correlationcoeff}) to the data, but the two models produce the same $chi^{2}$ goodness of fit metric (13 in this example), and thus the same $p$-value and $\sigma$ fit. 

While the $\chi^{2}$ goodness of fit test cannot show preference for one model over the other in this example, the PCC can provide more information. For model 1 (red line, Figure \ref{fig:correlationcoeff}) the PCC is -0.72, while model 2 (blue line, Figure \ref{fig:correlationcoeff}) has a PCC of 0.8, indicating that model 1 is very strongly \textit{negatively} correlated with the observations and model 2 is very strongly \textit{positively} correlated. In the physical example of a transmission spectrum as a function of wavelength, this may be interpreted as model 2 exhibiting absorption features when there is stronger absorption in the dataset. Therefore, one may conclude in this example that model 2 has the correct absorbers present, albeit too strongly, while model 1 is missing key features that would allow it to match the data.

\bibliography{sample631}{}
\bibliographystyle{aasjournal}



\end{document}